\documentclass[%
 reprint,
 amsmath,amssymb,
 aps,
pra,
floatfix,
]{revtex4-2}

\usepackage{graphicx}

\usepackage{csvsimple}
\usepackage{dcolumn}%
\usepackage{array} 

\usepackage{bm}
\usepackage[dvipsnames]{xcolor}

\usepackage{subcaption}
\usepackage{multirow}
\usepackage{times}
\usepackage{placeins}
\usepackage{hyperref}
\usepackage{booktabs}
\begin{document}
\hbadness=99999

\preprint{APS/123-QED}

\title{Phenomenology of renormalization group improved gravity from the kinematics of SPARC galaxies.}

\author{Esha Bhatia}
\email{b.esha@iitg.ac.in}
\author{Sayan Chakrabarti}
\email{sayan.chakrabarti@iitg.ac.in}
\author{Sovan Chakraborty}
\email{sovan@iitg.ac.in}
\affiliation{Department of Physics, Indian Institute of Technology, Guwahati 781039, India}


\date{\today}

\begin{abstract}
Renormalization Group correction to General Relativity (RGGR) proposes a logarithmic running of the gravitational coupling $\left(G\right)$, resulting 
in a modified description of gravity. Analogous to the proposal of dark matter, RGGR has the potential to explain the observed kinematics of the galaxies including the missing-mass problem. We, for the first time, based on the galaxy morphological types, investigate the dynamics of a diverse collection of galaxies present in the Spitzer Photometry for Accurate Rotation Curve (SPARC) catalog. We phenomenologically constrain the RGGR model parameter $\left(\bar\nu\right)$ along with the mass-to-light ratio for a sample of 100 SPARC galaxies, selected from four different morphological types, viz. early, spiral, late, and starburst. Our statistical analysis finds RGGR to fit the observed galaxy kinematics consistently. Additionally, we observe that apart from the galaxies where both RGGR and NFW fares equally well, RGGR is favored in more galaxies when compared to NFW. The constrained RGGR model parameter also supports the claim that it has a near-linear dependence on the galactic baryonic mass. From our morphology study, we find that the parameter $\bar\nu$ decreases from the early-type to the starburst galaxies. Finally, the renormalization group improved gravity is tested against the two established empirical relations for the SPARC catalog, viz., the Radial Acceleration Relation (RAR) and the Baryonic Tully Fisher relation (BTFR), both are found to satisfy consistently. Importantly, these results are 
are found to be nearly independent of the choice of the Bayesian priors on the model parameters. 

\end{abstract}

\maketitle


\section{Introduction} 
\label{sec:intro}

The proposal to modify the action of the general theory of relativity (GR) has been found to be successful \cite{Clifton:2011jh, Nojiri:2006ri} in mitigating the missing-mass problem \cite{Zwicky:1937zza, Rubin:1978kmz, Rubin:1980zd}. This leads to the introduction of additional components such as scalar, vector, or tensor fields in the action of gravity \cite{Clifton:2011jh, Nojiri:2006ri, Nojiri:2017ncd}. 
On the other hand, the dark matter (DM) explanation to these galactic kinematic issues is found to have wider implications in cosmology and astroparticle physics \cite{Perivolaropoulos:2021jda, Green:2021jrr, Khelashvili:2022ffq}. Although both modified gravity and DM scenarios are appealing, they suffer from certain challenges and shortcomings. 
On one hand, we have $\Lambda$CDM \cite{Perivolaropoulos:2021jda}, which has issues such as core-cusp problem \cite{de2010core}, coincidence problem \cite{Velten:2014nra}, missing satellite problem \cite{Klypin:1999uc, Kauffmann:1993gv} and many more \cite{Perivolaropoulos:2021jda, Weinberg:1988cp, Ostriker:2003qj, Moore:1994yx}. 
Similarly, any modified theory of gravity must demonstrate stability throughout all epochs of cosmic evolution without exhibiting any internal instabilities within its framework \cite{Clifton:2011jh, Tsujikawa:2010zza, DeFelice:2006pg}. Consequently, the detection of gravitational waves has led to the exclusion of several alternative theories that were inconsistent with the obtained phenomenological constraint on the mass of the graviton \cite{LIGOScientific:2016dsl, Yunes:2013dva}. Keeping all the past references in mind, our work in this paper discusses an alternative gravity scenario, which can explain the rotation curve for a large selection of galaxies. Specifically, we look into a model that studies the effect of the renormalization group (RG) running on the parameters under quantum field theory in curved spacetime. \\
\newline
Renormalization Group correction to General Relativity (RGGR) proposes running of the gravitational coupling ($G$) with the energy scale \cite{Reuter:2004nx, Shapiro:2004ch, Farina:2011me} and has 
been found to provide an alternative description to the galactic kinematic discrepancies \cite{Rodrigues:2009vf, Rodrigues:2012qm, Rodrigues:2012wk}. The previous studies on the RG effect showed that a logarithmic running of $G$ can phenomenologically explain the kinematics of galaxies \cite{Reuter:2004nx, Fabris:2012wg}. Additionally, the energy scale of the running parameter is related to the observable potential with the condition that it must satisfy the Tully-Fisher relation and regain the correct Newtonian limit on Solar system scales \cite{Shapiro:2000dz, Rodrigues:2014xka} as well as on astrophysical scales \cite{Domazet:2010bk}.  Similarly, earlier work on the RG study also accounts for the renormalization running of the cosmological constant ($\Lambda$) \cite{Shapiro:2000dz, Shapiro:2003ui}. However, due to the smallness of $\Lambda$, any modification arising out of the running of $\Lambda$ can be neglected on the galactic scales \cite{Rodrigues:2012qm, Rodrigues:2012wk, Rodrigues:2014xka}. The running of the gravitational constant results in a phenomenological parameter ($\bar\nu$) for the RGGR model. Studies on the kinematics of spiral and elliptical galaxies show that the $\bar{\nu}$ is of the order of $10^{-7}$ \cite{Rodrigues:2012qm, Rodrigues:2014xka, Oliveira2015TestingTA}. The parameter $\bar{\nu}$, which determines the strength of the running, has been found to have an almost linear relation with the baryonic mass of the galaxy and has been studied for a number of galaxies \cite{Rodrigues:2014xka}. In the following, we analyze the RGGR model with an even larger selection of galactic rotation curve (RC), as compiled in the Spitzer Photometry for Accurate Rotation Curve (SPARC) \cite{Lelli:2016zqa}. \\
\newline 
SPARC \cite{Lelli:2016zqa} is a collection of 175 rotationally supported galaxies that contain the radial variation of net circular velocity, measured at near-infrared region. This catalog has been extensively used for the phenomenological study of a large number of DM \cite{Ren:2018jpt, Khelashvili:2022ffq} and alternative gravity models \cite{ Naik:2019moz, deAlmeida:2018kwq, Green:2019cqm}. In addition to the net circular velocity, the catalog compiles the mass modeling of individual baryonic components, viz., disk, bulge, and gas of the galaxy. The SPARC encompasses galaxies belonging to different morphological types. This includes galaxies ranging from Hubble type $0-12$ that consist of Early, Spiral, Late-type, and Starburst galaxies. Our analysis of the RGGR model extends to all $4$ morphological types. We aim to study the dependence of the alternative gravity parameter on the galaxy morphology, and in this regard, we, for the first time, based on the galaxy morphological types, investigate the dynamics of a diverse collection of galaxies. 

SPARC imposes certain selection criteria on the quality of observational data and the inclination of the galaxies from our reference frame, which reduces the number of relevant galaxies to $153$. Additionally, we impose a $\chi^2$ cutoff to ensure that our analysis of the RGGR model shows a consistent fit with the observations. This further reduces the number of relevant galaxies to $100$. For each of these 100 galaxies, we constrain the RGGR parameter $\bar\nu$ from the observed RC data. The linear dependence of the parameter $\bar{\nu}$ for a wide range of masses in SPARC galaxies $(10^8-10^{11}~M_{\odot})$ is also determined in our analysis. The model parameters of the individual galaxies are statistically constrained using an algorithm based on the Bayesian technique known as Markov Chain Monte Carlo (MCMC) \cite{foreman2013emcee}. We also report the goodness of fit of individual galaxies by measuring the $\chi^2$ value to explain the consistency of the gravity model. \\
\newline
We extend the phenomenology of the RGGR model further to the well-known empirical relations, the Radial Acceleration Relation (RAR) and the Baryonic Tully-Fisher Relation (BTFR).  Independent of the underlying gravity model, an analytical relation exists between the observed and baryonic accelerations for all the SPARC galaxies. This relation is known as RAR \cite{mcgaugh2016radial}, which hints towards modified kinematics that governs the motion on galactic scales. Another such relation, known as the BTFR \cite{Tully:1977fu, Lelli:2019igz} signifies that the baryonic mass contained within all the galaxies present in the SPARC catalog has a power-law dependence on the flat part of the circular velocity ($V_f$) far from the galactic center.
 Using the best-fit values obtained from the RC analysis with the emcee sampler, we establish the above two relations of the SPARC catalog. As both the relations were evaluated from the observational data, having no model dependence, such relations became a consistency check for the gravity model used. Our elaborate analysis aims to check the consistency of the RGGR model for a rather large selection of rotationally supported galaxies to date. \\
\newline
Similar to the RGGR analysis, we look into the SPARC galaxies with an alternative DM scenario. In the DM paradigm, the net matter density of a galaxy, in addition to baryonic matter, has a contribution from DM \cite{Bertone:2004pz}. In our study of RC for the SPARC galaxies, we assume a particular well-known density profile obtained from the N-body simulation named Navarro-Frenk-White (NFW) \cite{Navarro:1995iw}. Here, similar to the RGGR analysis, we constrain the DM model parameter ($M_{200}$) using the Bayesian MCMC algorithm \cite{foreman2013emcee}. Furthermore, we compare the fit of the RC for the two profiles, i.e., RGGR and DM, by evaluating the Bayesian Inference Criteria (BIC) \cite{1978AnSta...6..461S} that helps to check the preferability of one model over the other. This analysis helps to check the consistency and the credibility of the RGGR model in comparison to an alternative DM (NFW) scenario.  \\
\newline
The layout of the paper is as follows. Section \ref{sec:rggr} gives a detailed account of the RGGR model. The next Section \ref{sec:gal} discusses in detail the galaxy catalog SPARC adopted for studying the alternative gravity model. The methodology and the computational packages used to sample the parameter space of the model are discussed in Section \ref{sec:method}. The following Section \ref{sec:res} and Section \ref{sec:conc} discuss the results obtained and the conclusion for our study, respectively.

\section{The models}
\label{sec:rggr}
\subsection{RGGR gravity model}

Renormalization Group correction to General Relativity is a quantum gravity model that studies the variation of the gravitational coupling parameter ($G$) with the energy scale of the Universe \cite{Reuter:2004nx, Farina:2011me, Rodrigues:2009vf}. The observations in the far infrared region established $G$ to be a constant in nature. However, this might not be true if one looks at GR as a field theory in curved spacetime. The beta function for the gravitational coupling constant $G$ has been studied to have a logarithmic dependence on the energy scale ($\mu$) of the Universe. From the dimensional argument, one unique choice for the beta function has been taken to be of the following form \cite{Farina:2011me,Rodrigues:2009vf}
\begin{equation}\label{bet}
    \beta=\mu\frac{dG^{-1}}{d\mu}=2\nu\frac{M^2_{planck}}{c\hbar}=2\nu G_0^{-1},
\end{equation}
where $\nu$ is a phenomenological parameter that is fixed from observational data, and $G_0$ is the bare value of the gravitational coupling parameter. The logarithmic dependence of $G$ with the energy scale as obtained from the solution of Eq.\ref{bet} becomes
\begin{equation}\label{gmu}
    G(\mu)=\frac{G_0}{1+\nu \ln\left(\frac{\mu^2}{\mu_0^2}\right)},
\end{equation}
here $\mu_0$ is the energy scale defined such that $G(\mu_0)=G_0$. As the variation in $G$ is small, the exact value of $\mu_0$ is inconsequential. For the far-infrared (IR) region, where the coupling parameter $G$ is known as a constant, the GR limit is regained by substituting $\nu=0$. Thus, the coupling parameter $G$ in the Einstein-Hilbert action of gravity follows the RG flow given in Eq.\ref{gmu}. For observations in galactic scales the variation in $\nu$ is of the order $10^{-7}$ \cite{Rodrigues:2014xka}. A similar $\beta$ function can also be defined for the cosmological constant ($\Lambda(\mu)$) present in the action of gravity. However, due to the negligible effect on the astrophysical scales, it is ignored from the RC analysis of the galaxies \cite{Rodrigues:2009vf}.\\
\newline
Additionally, to formulate a consistent theory with an observational basis, it is required to correlate the energy scale with some observable parameter. For example, on the cosmological scales, $\mu$ is shown to have a relation with the energy scale of the Universe i.e., Hubble constant ($H$) \cite{Fabris:2012wg}. Similarly, for the galactic scales, $\mu$ has a functional dependence on the potential energy of the system. The relation between the energy scale dependence of the parameter with the potential is defined such that it regains the correct Newtonian limit and satisfies the Tully Fisher relation. Thus, the functional form that satisfies the mentioned condition can be assumed to be \cite{Fabris:2012wg} 
\begin{equation}
    \frac{\mu}{\mu_0}=\left(\frac{\phi}{\phi_0}\right)^{\alpha},
\end{equation}
where $\alpha$ is a phenomenological parameter constrained from the observations. The analysis of galactic dynamics with RGGR \cite{Rodrigues:2012qm, Rodrigues:2014xka} suggests that the parameter $\nu\alpha\equiv\bar{\nu}$ follows a close to linear relation with the baryonic mass of the galaxy. Such a linear relation has been shown from the study of RC of spiral galaxies \cite{Rodrigues:2014xka} and from the fundamental plane of elliptical galaxies \cite{Rodrigues:2012qm}.
Solving for the weak-field limit, the effective circular velocity and acceleration take the following forms \cite{Rodrigues:2009vf, Rodrigues:2012qm}
\begin{equation}\label{vrgr}
    v_{RGGR}^2(r)=v_N^2(r)\left(1-\frac{c^2\bar{\nu}}{\phi_N(r)}\right),
\end{equation}
\begin{equation}\label{arggr}
    a_{RGGR}(r)\approx a_N(r)\left(1-\frac{c^2\bar{\nu}}{\phi_N(r)}\right),
\end{equation}
where $V_N(r)$ and $a_N(r)$ are the Newtonian contribution to the velocity and acceleration, respectively, $c$ is the speed of light.  
Thus, on the galactic scales, the net circular velocity and acceleration of an object, in addition to the Newtonian (baryonic) part, have the extra contribution due to the modified gravity. This extra contribution depends on the model parameter $\bar\nu$ and the potential $\phi_N(r)$. The potential $\phi_N(r)$ arises from all the baryonic contents of the galaxy and is crucial to evaluate the effect of RGGR gravity. This requires mass modeling, i.e., modeling the radial dependence of the baryonic mass of the galaxy. In general, the baryonic structure of the galaxy is composed of three components, i.e., disk, gas, and bulge. We assume a radial exponential profile \cite{Freeman:1970mx} for the disk and gas component. For the galaxies having bulge, the mass modeling is done assuming a spheroidal Hernquist profile \cite{Hernquist1990AnAM}. Thus, for the given mass profiles, one may estimate or constrain the phenomenological parameter $\bar{\nu}$ from observations. For example, rotation curve studies \cite{Rodrigues:2014xka} estimate $\bar{\nu}$ to be of the order of $10^{-7}$ to be consistent with the observations.

 Additionally, the constraints obtained on the solar system scales suggest the parameter $\bar{\nu}$ to be of the order of $10^{-17}$ \cite{Farina:2011me}.
The limit is much smaller than the galaxy rotation curve estimates. This, however, is expected as the parameter $\bar\nu$ is found to have a linear dependence on the baryonic mass of the system. In comparison to the spiral galaxies, the mass contained within the Solar system is about ten times less, thus making the $\bar{\nu} \sim 10^{-17}$ limit consistent. Similarly, in the case of Ultra-diffuse galaxies, whose masses are comparatively smaller than the rotationally supported galaxies, the evaluated $\bar \nu$ turns out to be $10^{-8}$ \cite{Bhatia:2023pts}. 
 In what follows, we analyze the consistency of the RGGR model for a large selection of galaxies from the SPARC catalog. These galaxies belonging to different morphologies with different characteristics give a measure of the model parameter's variation over the nature of the galaxies. \\ 
 \newline

 \subsection{NFW dark matter}
We also compare the RGGR model with an alternative scenario where we assume that the missing mass problem on the astrophysical scales can be resolved by the presence of DM. For this, we look into a known DM profile, i.e., Navarro-Frenk-White (NFW) profile \cite{Navarro:1995iw}, having the density $\rho(r)=\frac{\rho_s}{\frac{r}{r_s}\left(1+\frac{r}{r_s}\right)^2}$, where $\rho_s(r)$ and $r_s$ are the characteristic density and radius respectively. It has already been shown \cite{Rodrigues:2014xka} that a comparison of the RGGR model with the alternative gravity model MOND \cite{1983ApJ...270..365M} or NFW DM model \cite{Navarro:1995iw} shows a better or an equally consistent fit for the observed circular velocity of the spiral galaxies. Along the same line, in this work, we compare the RGGR model with an NFW DM profile for the SPARC galaxies and check the favourability of one over the other. Our analysis for RGGR versus DM looks into a scenario where the number of free parameters of the model is similar. Thus, in contrast to the two-parameter fit for the DM profile \cite{Rodrigues:2014xka}, we fit a single DM model parameter ($M_{200}$) by using a stellar-halo relation \cite{Dutton:2014xda}. 
The velocity contribution from the assumed NFW profile is \cite{Navarro:1995iw}
\begin{equation}\label{eq:nfw}
    v^2_{NFW}(r)=\frac{4\pi Gr_s^3\rho_s}{r}\left[-\frac{r}{r+r_s}+\log \left(1+\frac{r}{r_s}\right)\right].
\end{equation}
For the DM halo scenario, the contribution to the net circular velocity for a galaxy comes from the sum of the baryonic component ($v_N$) which for SPARC is expressed by Eq.\ref{eq:vn} and NFW profile ($v_{NFW}$) i.e.,
\begin{equation}\label{eq:vtnfw}
    v^2_{tot}(r)=v^2_N(r)+v^2_{NFW}(r),
\end{equation}
Both the model parameters of NFW ($\rho_s(r), ~ r_s$) can be expressed in terms of concentration parameter $c$ and virial mass $M_{200}$ \cite{Li:2020iib}. We assume an additional constraint which relates $c-M_{200}$ for a galaxy-sized halos as \cite{Dutton:2014xda},
\begin{equation}
    c(M_{200})=10^{0.905}\left(\frac{M_{200}}{10^{12}h^{-1}M_{\odot}}\right)^{-0.101},
\end{equation}
where h=0.671 \cite{Planck:2015fie}. This $c-M_{200}$ relation leaves only one parameter, i.e., $M_{200}$, that is constrained from the analysis of observed RC. A comparison of the RGGR model with an NFW DM halo helps to determine the favorability of the gravity model in comparison to a DM scenario. 

\section{Galaxy catlogue}
\label{sec:gal}
Spitzer Photometry for Accurate Rotation Curve (SPARC) contains the collection of $175$ rotationally supported galaxies measured at near-IR photometry \cite{Lelli:2016zqa}. The catalog contains the mass models of galaxies covering a broad range with varying luminosities, morphologies, rotation velocity, gas content, etc. \cite{Lelli:2016zqa}. The dominant contribution for a rotationally supported galaxy comes from the disk, which extends up to a few kiloparsecs. The additional baryonic contribution arises from the bulge (if present) superimposed at the disk's center and gas diffused throughout the galaxy. Depending upon the size of the bulge and structure of spiral arms present, the SPARC is majorly divided into $4$ morphological types. The different morphologies are represented via Hubble type ($H$), which for the catalog ranges between $H: 0-12$ \cite{mihalas1981galactic}. The Hubble type is a representation of the evolutionary stage of the galaxy. In general, the size of the bulge and tightness of spiral arms for a rotationally supported galaxy reduces as we move up in the Hubble type.  The first Hubble category ranging from $H: 0-2$ belongs to the Early type, which contains galaxies of type S0, Sa, and Sab, respectively. Such galaxies are distinguished by the presence of a prominent bulge at the center and tightly woven spiral arms. The second category consists of Spiral galaxies ($H: 3-6$) where the features such as bulge and tightness of spiral arms start decreasing. The next two categories include Late-type ($H: 7-9$) and Starburst ($H: 10-12$). These galaxies contain almost no visible bulge present at the center of the galaxy. The last category i.e., Starburst galaxies are known to have no spiral structure in the outer parts and are characterized by their diffused shape. The Starburst galaxies are also known to have a higher star formation rate. Thus, this catalog represents a wide spectrum of galaxies and  makes a versatile ground to test our alternative gravity model.\\
\newline
\textbf{\textit{Rotation Curve:}} Given the radial variation of the velocity for individual baryonic components of a galaxy, the total Newtonian contribution can be written as:
\begin{equation}\label{eq:vn}
    v_{N}^2(r)=\gamma_d~ v_{disk}^2(r)+\gamma_b v_{bulge}^2(r)+|v_{gas}(r)|~v_{gas}(r);
\end{equation}
where $v_{disk}(r)$, $v_{bulge}(r)$ and $v_{gas}(r)$ represent the disk, bulge, and gas velocity components for a particular galaxy in the SPARC catalog. The radial variation of the individual baryonic component for each rotationally supported galaxy is provided within the SPARC. The catalog additionally contains the accurate HI and H$\alpha$ measurement of the total circular velocity $v_{obs}(r)$, along with the error bars $\sigma(r)$ for every galaxy. Also, the two baryonic components are scaled by a factor $\gamma_d$ and $\gamma_b$, which measure the mass-to-light ratio for the disk and bulge part, respectively. The data from the catalog clearly shows that the net contribution of velocity coming from the baryonic component is insufficient to explain the observed velocity, $v_{obs}(r)$ of the galaxy. This leaves room for the additional components, such as DM or MOG which can be added to the Newtonian part to explain the overall rotational stability of the galaxy.  \\
\newline 
Under the assumption that the gravity on the galactic scale is RGGR, the modified kinematics of objects within the galaxy is expressed as given in Eq.\ref{vrgr}. The equation shows that the net circular velocity has an additional contribution from the RGGR gravity.
Therefore, comparing the total observed circular velocity from the SPARC catalog with the RGGR velocity form (Eq.\ref{vrgr}) helps to check the consistency of the gravity model. This requires the variation of the model parameter of the theory within the allowed parameter space. This analysis aims to look for the parameters that give a consistent fit to the observed RC of the galaxy sample. The best-fit values obtained in the analysis are also used to construct the empirical relation RAR and BTFR.  \\
\newline
\textbf{\textit{RAR:}} The analysis of SPARC data shows that the total baryonic acceleration $(a_{bar})$ cannot explain the net observed acceleration $(a_{obs})$ and follow a certain analytical relation known as RAR \cite{mcgaugh2016radial}. This analysis has been found to be true for the 153 galaxies of the catalog irrespective of their morphological types, thus indicating a new dynamical law governing galaxy kinematics. The empirical RAR relationship obtained from SPARC is defined as \cite{mcgaugh2016radial}
\begin{equation}\label{rar}
    a_{obs}(R)=\frac{a_{bar}(R)}{1-\exp(-\sqrt{a_{bar}(R)/a_*})},
\end{equation}
where $a_*$ is the acceleration scale parameter and has best-fit value $a_*=1.2\times 10^{-10}$ ms$^{-2}$. 
For a rotationally supported galaxy, the net centripetal acceleration is defined in terms of observed velocity in SPARC catalog
\begin{equation}\label{acc}
    a_{obs}(R)=\frac{v_{obs}^2(R)}{R}=\left|\frac{\partial \phi_{tot}(R)}{\partial R}\right|,
\end{equation}
where $\phi_{tot}$ is the total potential i.e., total force per unit mass acting on a point particle. Similarly, $a_{bar}(R)$ is the linear sum of the acceleration for different baryonic components (disk, gas, and bulge) within the galaxy and can be estimated from the SPARC data using
\begin{equation}\label{bar}
\begin{split}
    a_{\text{bar}}(R) &= \frac{v_{N}^2(R)}{R} \\
    &= \frac{\gamma_d~v_{\text{disk}}^2(R) + \gamma_b~ v_{\text{bulge}}^2(R) + |v_{\text{gas}}(R)|~v_{\text{gas}}(R)}{R} \\
    &= \left|\frac{\partial \phi_{\text{bar}}(R)}{\partial R}\right|.
\end{split}
\end{equation}
Both the net circular velocity ($v_{obs}(r)$) and individual baryonic components mentioned in Eq.\ref{acc} and \ref{bar} are known observationally in SPARC. This makes the relation an empirical one as it assumes no prior knowledge about the DM or MOG model and is purely from the observational data in SPARC.  \\
\newline
For our analysis, we aim to probe RAR in the context of the RGGR gravity. The analytical expression for the net acceleration (Eq.\ref{arggr}) in RGGR includes a $\bar{\nu}$ dependent extra component to the Newtonian part. As explained in the previous discussion in the context of RC, the best-fit value of this $\bar\nu$ parameter for each galaxy can be estimated from the circular velocity fitting with the SPARC data. Thus, the consistency of the RGGR predicted net acceleration ($a_{RGGR}$) with respect to the $a_{obs}(R)$ can be probed by the RAR. In the following, we compare the observed RAR with the same empirical relation Eq.\ref{rar} in the context of the RGGR model for each data point of the qualifying SPARC galaxies. \\
\newline
\textbf{\textit{BTFR:}} SPARC also shows a tight correlation between the dynamics of the galaxy with the baryonic distribution. BTFR suggests that the stellar mass of the galaxy has a power law dependence on the flat part of the circular velocity, 
\begin{equation}\label{btfr}
    M_{bar}=A V_f^x.
\end{equation}
Here $M_{bar}$ refers to the baryonic mass contained within the galaxy and $V_f$ is the velocity measured along the flat part of the RC. 
The optimal value of the free parameters ($A, x$) are obtained from the study of the SPARC catalog, and are estimated to be $A=50$ M$_{\odot}$km$^{-4}$s$^4$, $x=4$ \cite{McGaugh:2005qe}. Similar to RAR, BTFR is also known to be empirical in nature as it assumes no underlying gravity model and is strictly obtained from the SPARC data. For every galaxy in SPARC, the baryonic mass is the linear sum of disk, bulge, and gas components.  For the RGGR model, the total circular velocity is estimated by $v_{RGGR}$ (Eq.\ref{vrgr}) and is dependent on the free parameters ($\gamma_d$, $\gamma_b$, $\bar{\nu}$) which are already constrained using the observed RC of the galaxy. Thus, the $V_f$ for the RGGR will also get modified. The relation between the $M_{bar}$ and the RGGR predicted flat velocity for each qualifying SPARC galaxy should follow observed BTFR. This additional check for RGGR is crucial for the gravity model to remain phenomenologically consistent.  

\section{Methodology}
\label{sec:method}
Using the RC data for the qualifying SPARC galaxies, we fit the model parameters considering RGGR as the underlying gravity model. These free parameters are constrained using the publicly available $emcee$ package \cite{foreman2013emcee} in PYTHON. This package works on the principle of Markov Chain Monte Carlo (MCMC), which samples the posterior distribution of the free model parameters. For a given observational data set $\mathcal{D}$, the posterior probability $\mathcal{P}$($\theta$|$\mathcal{D}$) for unknown set of parameters $\theta$ is defined as
\begin{equation}\label{sampler}
    \mathcal{P}(\theta|\mathcal{D})\propto \mathcal{L}(D|\theta)\pi(\theta),
\end{equation}
where $\mathcal{L}(D|\theta)$ is the likelihood which determines the probability of a data for a given model with free parameters, and $\pi(\theta)$ represents the priors imposed on the free parameters. Assuming that the errors on the observed circular velocity follow a Gaussian distribution, the likelihood for each galaxy is written as
\begin{equation}\label{eq:like}
\begin{split}
    \mathcal{L}_g &= (2\pi)^{-n/2} \left\{ \prod_{i=1}^n \sigma(r_i)^{-1} \right\} \\
    &\quad \times \exp \left\{ -\frac{1}{2} \sum_{i=1}^n \left( \frac{v_{g,\text{obs}}(r_i) - v_{\text{tot}}(r_i,\vec{\theta})}{\sigma(r_i)} \right)^2 \right\},
\end{split}
\end{equation}

here, $n$ represents the number of observational data points over which the likelihood is summed. Also, $v_{obs}(r)$ and $\sigma(r)$ are the total circular velocity and error on the observations defined at each radial point within the SPARC catalog. The analytical velocity $v_{tot}(r,\vec{\theta})$ computed at a certain radius for a given set of free parameters is model-dependent. 
We fit both the RGGR and the NFW DM models. The net circular velocities for RGGR and NFW in terms of their respective free parameters are expressed by Eq.\ref{vrgr} and  Eq.\ref{eq:vtnfw}, respectively. These free parameters in velocity $v_{tot}(r,\vec{\theta})$ are phenomenologically constrained by comparing with the observational circular velocity. For both the models, i.e., RGGR and NFW, the normalization mass-to-light factors ($\gamma$) introduced for the baryonic components, i.e., the disk ($\gamma_d$) and the bulge ($\gamma_b$) of the galaxy are common. Additionally, the RGGR model has a mass-dependent phenomenological parameter $\bar\nu$. Similarly, the NFW model has a singular free parameter $M_{200}$. Thus, in the alternative gravity scenario $\vec{\theta}$ is composed of \{$\gamma_d$, $\gamma_b$, $\bar{\nu}$\} which are to be estimated independently for each galaxy. For the case of NFW $\vec{\theta}$ is made up of \{$\gamma_d$, $\gamma_b$, $M_{200}$\} and are aagin constrained for each galaxy.
\newline
Regarding the priors $\pi(\vec{\theta})$, we assume both flat and Gaussian priors for the model parameters. In case of the flat priors, the mass-to-light ratio for both disk ($\gamma_d$) and bulge ($\gamma_b$) are assumed to have no radial dependence on the galaxy and are varied in the range [$0.3$, $0.8$] \cite{deAlmeida:2018kwq, schombert2014stellar, Meidt:2014mqa}. In the case of RGGR, for the mass-dependent parameter $\bar{\nu}$, a wide range of parameter space is looked into and is varied within $10^{-9}$ $\le$ $\bar{\nu}$ $\le$ $10^{-6}$. This range is motivated by the previous analysis of spiral and elliptical galaxies where $\bar{\nu}$ is found to vary in order of $10^{-7}$ \cite{Rodrigues:2014xka}. For the DM scenario, the NFW parameter $M_{200}$ is bound to lie within the range $10^9<M_{200}/M_{\odot}<10^{14}$ \cite{deAlmeida:2018kwq}. For the Gaussian priors both the parameters $\gamma_d$ and $\gamma_b$ are considered to have a mean value of $0.5$ and standard deviation of $0.1$. The mean and standard deviation for $\log(\bar\nu/10^{-7})$ and $\log(M_{200}/ M_{\odot})$ are taken to be ($-0.3,~0.1$) and ($10.69,~0.1$),  respectively. To ensure the convergence of the chain, we run the sampler for a sufficient number of steps such that the acceptance fraction lies within the range $0.2-0.5$ \cite{foreman2013emcee}. Additionally, by estimating the autocorrelation time ($\tau$) for each galaxy, we discard $\tau$ number of steps as burn-in before performing posterior analysis. We run a sufficient number of steps, i.e., $50~\tau$ as specified in $emcee$ \cite{foreman2013emcee}, to achieve convergence. \\

Additionally, to quantify the preference of one model over the other, we consider the Bayesian Inference Criteria (BIC), which is a simple approximation for evidence and is defined as \cite{1978AnSta...6..461S}
\begin{equation}\label{eq:bic}
    BIC=-2\log \mathcal{L}_{max}(D|\vec{\theta})+2k\log(n).
\end{equation}
Here $k$ is the number of parameters for a model which is the same, i.e., $3$ parameters, for both the RGGR and NFW models considered here. Also, $n$ represents the number of data points in each galaxy. To compare the two models, i.e., RGGR and NFW, we evaluate,
\begin{equation}
    \Delta BIC=BIC_{NFW}-BIC_{RGGR}.
\end{equation}
The measure of $\Delta BIC$, if less than $2$, implies inconclusive preference between the two models. Similarly, the value of $\Delta BIC$ that is between $2-6$ implies a positive inclination towards RGGR. Values of $\Delta BIC$ greater than $6$ are considered to have a strong inclination toward the RGGR model.
\newline

Certain selection criteria are adopted before choosing a galaxy from the SPARC catalog. This includes galaxies with asymmetrical RC and poor quality factor ($Q$), i.e., $Q > 2$. We also exempt face-on galaxies having $i<30^{\circ}$ from our analysis \cite{Lelli:2016zqa}. To express definite $\chi^2_{red}$, we assume galaxies with data points greater than $4$. Additionally, to ensure that we have a consistent RC fit with the RGGR, we reject galaxies whose goodness of fit with the gravity model, i.e., $\chi^2_{red} \ge 6$. The last quality cut is imposed manually to ensure a consistent fit for the model with the observations.\\
\begin{table*}[ht!] 
    \caption{The best-fit parameters obtained from the emcee sampler. Corresponding to the RGGR gravity model, for every galaxy, 3 parameters ($\gamma_d$, $\gamma_b$, $\bar{\nu}$) are statistically constrained from the observed circular velocity. Column 2 of the table denotes the galaxy morphology types, abbreviated as E: Early, S: Spiral, L: Late-type, and SB: Starburst. In the table, the representation for the free parameter $\bar{\nu^*}=\bar \nu \times 10^{-7}$ and $M^*_{200}=M_{200} \times 10^{11} M_{\odot}$.}
    \label{par_sparc}
    \centering
    \begin{tabular}{|c|c|c|c|c|c|c|c|c|c|}
        \hline
        \textbf{Galaxy} & \textbf{Type} & \multicolumn{2}{c|}{\textbf{$\gamma_d$}} & \multicolumn{2}{c|}{\textbf{$\gamma_b$}} & \textbf{$\bar{\nu}^*$} & \textbf{$M^*_{200}$} & \multicolumn{2}{c|}{\textbf{$\chi^2_{red}$}} \\
        \hline
        \textbf{Name} &  & $\gamma_d$ (RGGR) & $\gamma_d$ (NFW) & $\gamma_b$ (RGGR) & $\gamma_b$ (NFW) &  &  & $\chi^2_{red}$ (RGGR) & $\chi^2_{red}$ (NFW) \\
        \hline
        \begin{filecontents*}{sample.csv}
Name,Type,gamadrgr,gamadnfw,gamabrgr,gamabnfw,nubar,mpar,chirgr,chinfw
NGC7814,E,0.53,0.79,0.28,0.66,7.59,24.13,0.99,1.12
NGC4138,E,0.58,0.79,0.5,0.31,1.45,3.32,0.61,0.7
UGC02487,E,0.8,0.8,0.69,0.8,7.59,60.94,2.57,17.1
UGC06614,E,0.52,0.31,0.33,0.31,3.47,12.89,1.08,0.54
UGC03546,E,0.59,0.65,0.35,0.37,2.14,8.93,0.91,0.87
UGC03205,E,0.55,0.72,0.95,0.8,3.47,10.61,4.09,4.13
UGC08699,E,0.69,0.78,0.52,0.67,1.66,7.81,1.41,0.64
UGC11914,E,0.3,0.63,0.58,0.8,8.32,100,1.68,4.31
UGC06786,E,0.4,0.8,0.55,0.75,4.37,26.14,2.29,2.2
NGC1090,S,0.54,0.47,0.5,0.5,1.66,4.88,2.86,2.2
NGC2683,S,0.58,0.7,0.5,0.36,1.82,4.18,0.71,0.84
NGC2841,S,0.73,0.8,0.86,0.8,5.62,40.6,1.73,6.08
NGC3769,S,0.36,0.36,0.5,0.5,1.35,2.61,0.99,0.56
NGC2955,S,0.34,0.31,0.58,0.73,4.37,23.23,4.97,3.25
NGC4013,S,0.2,0.42,0.64,0.76,2.69,7.33,1.64,1.35
NGC0801,S,0.57,0.52,-,-,1.86,7.87,5.81,6.5
NGC5005,S,0.66,0.5,0.45,0.49,0.66,38.39,0.39,0.05
NGC6195,S,0.31,0.31,0.55,0.64,4.47,15.27,1.95,2.21
NGC7331,S,0.3,0.36,0.46,0.36,3.63,20.07,2,0.75
UGC12506,S,0.79,0.79,-,-,0,19.69,3.76,1.58
UGC11455,S,0.41,0.4,-,-,5.01,31.32,4.16,5.51
UGC09037,S,0.3,0.3,-,-,2.24,3.09,3.18,5.81
UGC07151,S,0.78,0.65,-,-,0.37,0.43,2.11,2.23
UGC06983,S,0.79,0.73,-,-,1.02,1.96,1.99,0.58
NGC4559,S,0.41,0.33,-,-,0.69,2.31,1.03,0.22
NGC4100,S,0.63,0.72,-,-,1.91,5.09,2.12,1.16
NGC4088,S,0.31,0.31,-,-,1.62,4.86,1.11,0.5
NGC4085,S,0.34,0.31,0.5,0.53,1.1,7.28,3.18,3.51
NGC3972,S,0.62,0.33,-,-,1.32,7.95,2.06,0.99
NGC3893,S,0.43,0.5,-,-,2.24,11.6,0.64,1
NGC2998,S,0.66,0.59,-,-,2.57,9.86,2.48,2.11
NGC0100,S,0.25,0.32,-,-,1.2,0.9,0.37,2.3
ESO079*,S,0.77,0.32,-,-,2.19,14.22,3.85,4.46
NGC0289,S,0.64,0.54,-,-,2.29,6.46,1.89,1.78
NGC2976,S,0.56,0.76,-,-,0.87,0.02,0.51,1.68
NGC3521,S,0.45,0.51,-,-,2.14,17.03,0.76,0.29
NGC3949,S,0.5,0.39,-,-,0.76,14.01,0.85,0.19
NGC3953,S,0.76,0.69,-,-,0.65,2.11,0.56,0.34
NGC3992,S,0.75,0.78,-,-,3.63,14.08,1.9,0.63
NGC4051,S,0.6,0.48,-,-,0.44,1.67,1.41,0.62
NGC4183,S,0.88,0.68,-,-,0.81,1.39,1.28,0.16
NGC4157,S,0.34,0.39,0.49,0.42,2.24,7.93,0.84,0.44
NGC0891,S,0.26,0.35,0.4,0.5,3.31,25.01,5.26,5.33
NGC5055,S,0.36,0.38,-,-,1.91,8,2.82,7.54
UGC02885,S,0.57,0.45,0.75,0.79,5.5,38.97,0.98,1.54
F583-4,S,0.53,0.4,-,-,0.5,0.32,0.26,0.69
NGC6946,S,0.44,0.47,0.46,0.5,1.74,5.62,1.5,1.5
NGC6503,S,0.41,0.54,-,-,1.17,2.3,5.98,2.57
NGC4010,L,0.43,0.32,-,-,1.29,3.51,1.57,2.67
NGC0300,L,0.65,0.32,-,-,0.76,1.48,0.78,0.97
UGC04278,L,0.9,0.33,-,-,0.69,0.7,3.05,3.82
UGC05721,L,0.72,0.78,-,-,0.72,1.73,2.32,3.31
NGC7793,L,0.62,0.57,-,-,0.55,1.24,0.64,0.84
UGC06446,L,0.79,0.73,-,-,0.71,0.94,2.86,0.35
UGC07603,L,0.41,0.4,-,-,0.47,0.7,0.66,1.23
UGC06930,L,0.74,0.54,-,-,0.83,1.45,1.65,0.23
\end{filecontents*}
\csvreader[head to column names]{sample.csv}{}%
        {%
            \Name & \Type & \gamadrgr & \gamadnfw & \gamabrgr & \gamabnfw & \nubar & \mpar & \chirgr & \chinfw \\
            \hline
        }
    \end{tabular}
\end{table*}

\begin{table*}[ht!] 
    \caption*{Continued from the previous page}
    \centering
    \begin{tabular}{|c|c|c|c|c|c|c|c|c|c|}
        \hline
        \textbf{Galaxy} & \textbf{Type} & \multicolumn{2}{c|}{\textbf{$\gamma_d$}} & \multicolumn{2}{c|}{\textbf{$\gamma_b$}} & \textbf{$\bar{\nu}^*$} & \textbf{$M^*_{200}$} & \multicolumn{2}{c|}{\textbf{$\chi^2_{red}$}} \\
        \hline
        \textbf{Name} &  & $\gamma_d$ (RGGR) & $\gamma_d$ (NFW) & $\gamma_b$ (RGGR) & $\gamma_b$ (NFW) &  &  & $\chi^2_{red}$ (RGGR) & $\chi^2_{red}$ (NFW) \\
        \hline
        \begin{filecontents*}{sample2.csv}
Name,Type,gamadrgr,gamadnfw,gamabrgr,gamabnfw,nubar,mpar,chirgr,chinfw
UGC08550,L,0.85,0.67,-,-,0.3,0.27,3.44,0.55
UGC12732,L,0.79,0.35,-,-,0.81,0.9,2.53,0.67
UGC11557,L,0.37,0.35,-,-,0.37,0.15,0.87,2.3
UGC08490,L,0.8,0.79,-,-,0.6,0.87,0.51,1.39
UGC07089,L,0.48,0.36,-,-,0.49,0.26,0.14,2.63
UGC07261,L,0.59,0.63,-,-,0.63,0.57,1.06,0.12
UGC07323,L,0.55,0.35,-,-,0.59,0.57,0.22,2.86
UGC07399,L,0.69,0.76,-,-,1.23,7.96,2.82,3.01
UGC06399,L,0.63,0.38,-,-,0.72,1.12,0.71,0.94
UGC06917,L,0.69,0.35,-,-,0.98,2.51,1.76,0.61
UGC05986,L,0.43,0.54,-,-,1.7,4.8,1.08,5.1
UGC04499,L,0.7,0.34,-,-,0.41,0.44,2.12,1.39
NGC3109,L,0.5,0.35,-,-,0.54,0.15,0.89,36.92
NGC0055,L,0.35,0.31,-,-,0.94,0.38,1.18,12.06
ESO116*,L,0.63,0.42,-,-,0.91,3.26,1.91,2.21
IC2574,L,0.42,0.8,-,-,0.44,0.02,2.73,196.86
F568-3,L,0.7,0.49,-,-,1.2,0.55,2.07,8.54
F568-V1,L,0.79,0.65,-,-,2.04,6.02,2.37,0.57
F571-V1,L,0.56,0.53,-,-,0.55,0.61,0.55,0.86
F574-1,L,0.82,0.49,-,-,1.17,1.49,3.46,1.67
UGC05750,L,0.65,0.61,-,-,0.54,0.15,1.27,3.05
UGCA442,L,0.52,0.46,-,-,0.28,0.24,2.68,9.33
UGC10310,L,0.7,0.63,-,-,0.94,0.33,2.34,0.49
UGC07524,L,0.79,0.38,-,-,0.55,0.51,3.74,2.62
UGC07125,L,0.83,0.34,-,-,0.32,0.18,2.6,3.17
UGC00891,L,0.35,0.34,-,-,15.14,0.27,4.08,24.08
UGC08837,SB,0.44,0.4,-,-,0.18,0.01,0.5,10.56
UGC07559,SB,0.46,0.39,-,-,0.13,0.02,0.39,1.61
UGC07690,SB,0.59,0.71,-,-,0.28,0.17,0.65,0.46
UGC07866,SB,0.49,0.38,-,-,0.13,0.11,0.36,4.09
UGC06923,SB,0.49,0.38,-,-,0.49,0.97,0.77,0.82
UGC05414,SB,0.49,0.37,-,-,0.36,0.19,0.07,4.45
UGC05829,SB,0.74,0.44,-,-,0.37,0.16,2.27,1.32
NGC4068,SB,0.42,0.39,-,-,0.13,0.02,0.29,3.41
NGC3741,SB,0.65,0.37,-,-,0.3,0.11,2.7,4.72
ESO444*,SB,0.56,0.44,-,-,0.32,0.75,1.21,0.68
UGC05918,SB,0.6,0.37,-,-,0.33,0.09,1.93,0.46
F565-V2,SB,0.55,0.44,-,-,0.6,0.68,1.04,1.23
UGC09992,SB,0.48,0.46,-,-,0.16,0.03,0.56,0.22
NGC2915,SB,0.68,0.37,-,-,2.24,1.36,2.09,1.09
UGC05005,SB,0.55,0.41,-,-,0.62,0.69,0.23,2.42
UGC07577,SB,0.4,0.36,-,-,0.05,0.01,0.92,7.92
KK98-251,SB,0.52,0.47,-,-,0.09,0.01,0.17,7.75
UGCA444,SB,0.57,0.44,-,-,0.09,0.03,0.78,0.77
\end{filecontents*}
\csvreader[head to column names]{sample2.csv}{}%
        {%
            \Name & \Type & \gamadrgr & \gamadnfw & \gamabrgr & \gamabnfw & \nubar & \mpar & \chirgr & \chinfw \\
            \hline
        }
    \end{tabular}
\end{table*}
\section{Results}
\label{sec:res}
In the following, we analyze the consistency of the RGGR framework with the observational circular velocity for our selection of SPARC galaxies, we also draw a comparison of the RGGR model with NFW DM scenario. The RC analysis is done for galaxies belonging to different morphological types. The free model parameters are constrained using the emcee sampler, which scans the parameter space to evaluate the best-fit values. For this, we look into the rotation curves of individual galaxies present within SPARC. We also study the mass-dependent nature of the phenomenological parameter $\bar{\nu}$. These best-fit values must also satisfy the fundamental relationships RAR and BTFR, hence we compare our results for the RGGR model with both these relationships. 
For galaxies consistent with the RGGR model, we alternatively look into their kinematics in a DM-dominated case. We employ a similar Bayesian technique to constrain the model parameter $M_{200}$ for the NFW profile. To measure the favorability of the RGGR model over the DM NFW profile, we also report the $\Delta BIC$ value evaluated for each galaxy.

\subsection{Fit to the observed Rotation Curve}
\label{sec:rc}

Fitting the RC of each SPARC galaxy with the RGGR gravity model constrains the free parameters. It includes two mass modeling parameters defined for the baryonic component of a galaxy i.e., $\gamma_d$ and  $\gamma_b$. The additional model parameter, i.e., $\bar{\nu}$, comes from the choice of RGGR gravity and has been found to have an almost linear dependence on the luminous mass of the galaxy. We analyze the behavior of the phenomenological parameter $\bar{\nu}$ with the baryonic mass for the large sample of SPARC galaxies. In particular, we study the consistency of the RGGR-governed model for all four morphological types of the galaxies present in SPARC. This compares the phenomenological consistency of the RGGR gravity for all the different galaxy types. A similar RC analysis where instead of modified gravity, galaxies are assumed to be DM dominated with the radial density profile having the NFW form is also looked into. For this scenario, in addition to the mass model parameters $\gamma_d$ and $\gamma_b$, we fit $M_{200}$, which comes from the choice of NFW halo.

\begin{figure*}[t]
    \centering
\begin{subfigure}[t]{0.45\textwidth}
\includegraphics[width=\linewidth,height=0.9\linewidth]{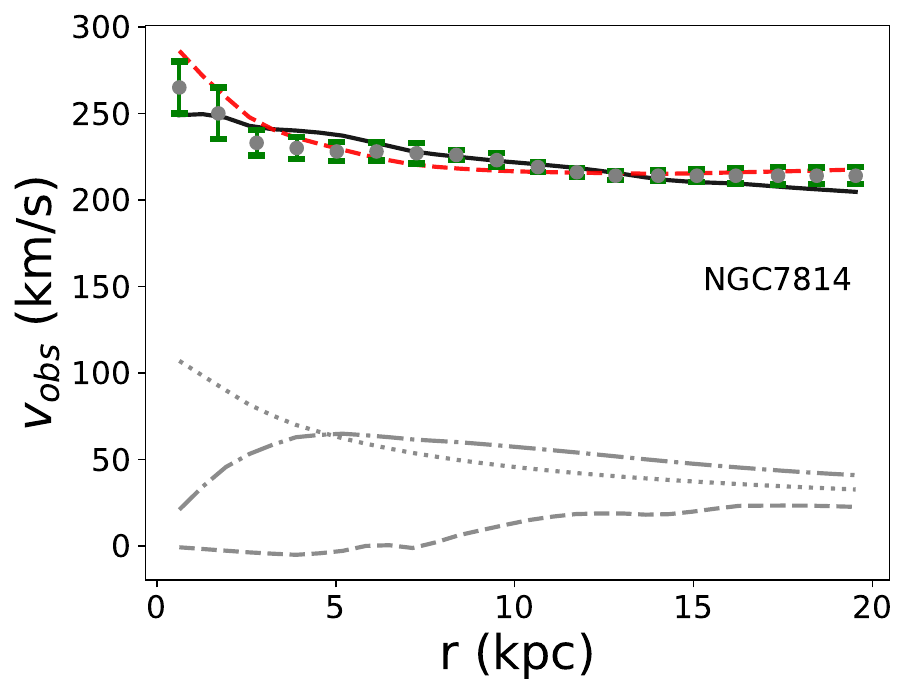} 

\end{subfigure}
\begin{subfigure}[t]{0.45\textwidth}
\includegraphics[width=\linewidth,height=0.9\linewidth]{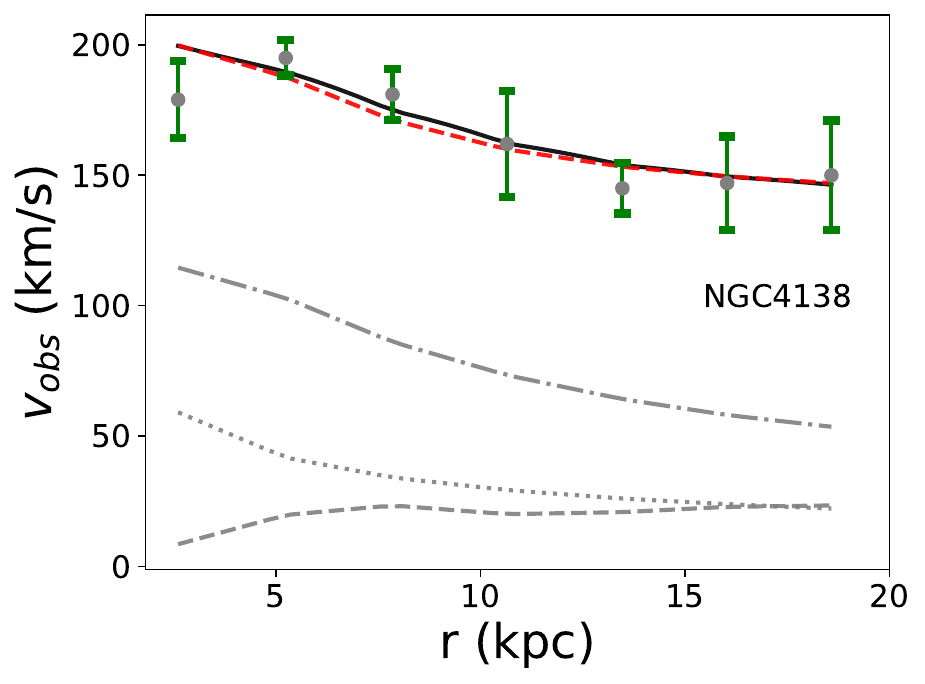} 
\end{subfigure}
\begin{subfigure}[t]{0.45\textwidth}
\includegraphics[width=\linewidth,height=0.9\linewidth]{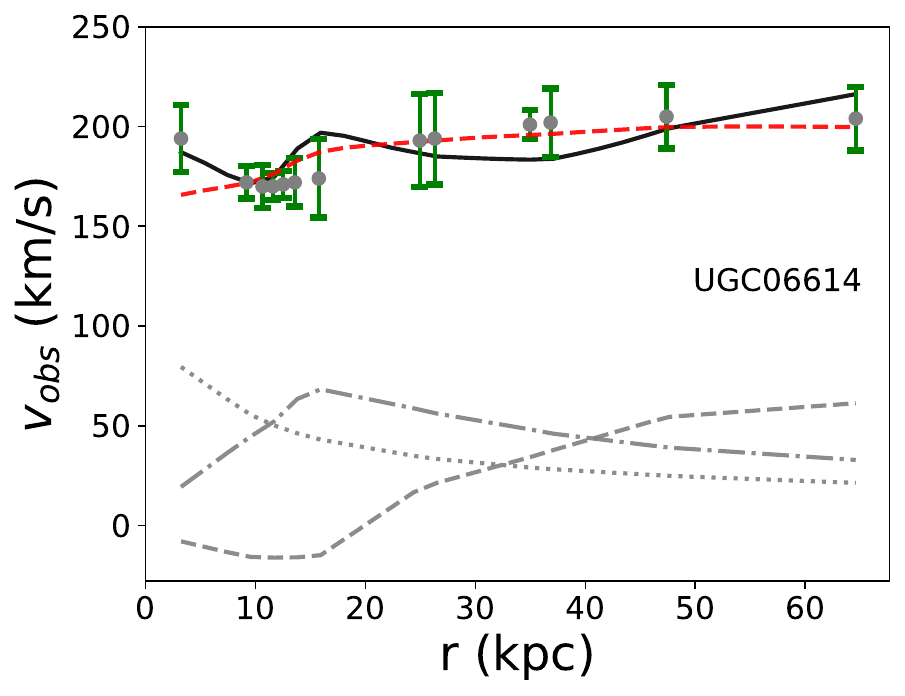} 
\end{subfigure}
\begin{subfigure}[t]{0.45\textwidth}
\includegraphics[width=\linewidth,height=0.9\linewidth]{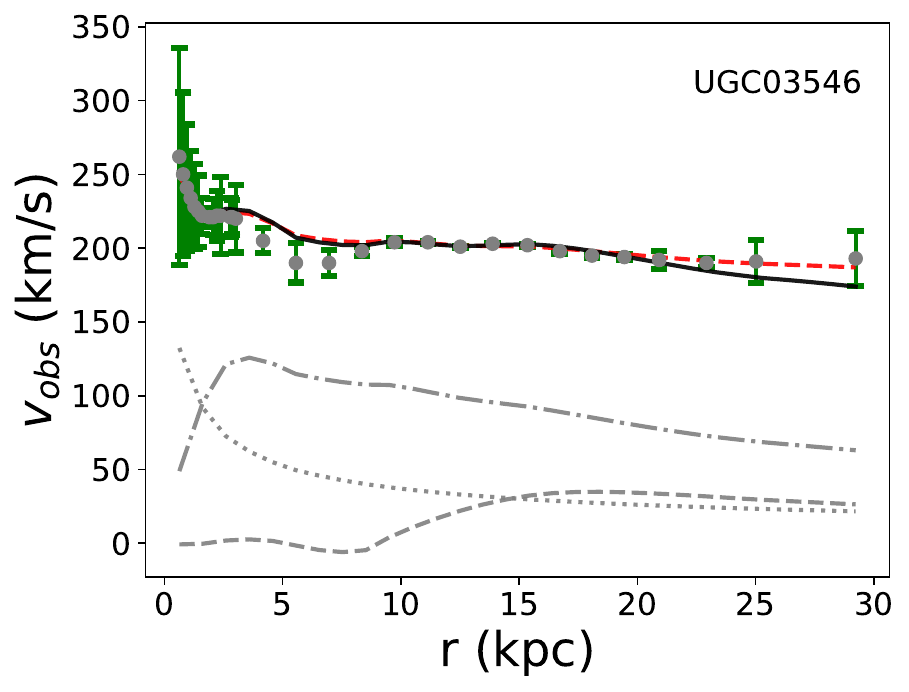} 
\end{subfigure}
\caption{For Early-type galaxies the RC for $4$ specimen galaxies are shown in the four panels above. The black solid line is the circular velocity when RGGR gravity contributes in addition to the baryonic part. The black dots with green error bars are the circular velocity data obtained from SPARC \cite{Lelli:2016zqa}. The red dashed line is where the net velocity contribution is evaluated in an NFW paradigm.  The dashed grey line represents the gaseous component of the galaxy. The dot-dashed line shows the variation of disk velocity within the galaxy, and the bulge part is plotted via the dotted line. }
\label{Fig:early}
\end{figure*}

\paragraph{\textbf{Early-type galaxies}}
\hfill \break
The first category belongs to the early-type galaxies, which can be identified from the presence of a bulge at the center and tight, indistinguishable spiral arms in the outer parts of the disk. From the whole set of early-type, we analyze $9$ galaxies that satisfy the selection criteria. Out of these $9$ galaxies, we illustrate the RC fitting for $4$ of them in Fig.\ref{Fig:early}. This includes NGC7814 shown in the upper left panel, NGC4138 in the upper right, UGC06614 in the lower left, and UGC03546 in the lower right panel of Fig.\ref{Fig:early}. For each galaxy shown, the grey dots with green error bars represent the observational total circular velocity traced by the HI component within the galaxy. We also plot the individual baryonic components, i.e., disk, bulge, and gas, for each galaxy using the gray lines. The dashed gray line represents the radial variation of the gas velocity. The disk and bulge components of the galaxy scaled by their respective mass-to-light ratio are shown via. the dashed-dotted and dotted lines, respectively.

In particular, the first panel at the top left shows the RC for NGC7814. Under the assumption of flat priors, the constraints on the free parameters ($\gamma_d, \gamma_b $ and  $\bar{\nu} \times 10^7$) obtained from our analysis give $0.53$, $0.31$ and $7.23$, respectively. However, when the priors are taken to be Gaussian we see almost no change in the best-fit value. As an example, for the case of NGC7814, when priors are assumed to be Gaussian ($\gamma_d, \gamma_b $ and  $\bar{\nu} \times 10^7$) are found to be $0.53$, $0.28$ and $7.53$ respectively. The radial variation of the circular velocity (Eq.\ref{vrgr}) in the RGGR model with these best-fit parameters is plotted using a solid black line. The obtained value of $\chi^2_{red}$ for the galaxy is $1.18$, indicating a good fit to the observational data of NGC7814. 

A similar RC analysis of the $9$ early-type galaxies in the presence of DM with the NFW profile is also looked into. The statistical analysis to constrain the model parameters of the DM halo gives the best-fit values evaluated in the case of flat-priors for $\gamma_d$, $\gamma_b$, and $M_{200}\times 10^{11} M_{\odot}$ as $0.79$, $0.66$ and $24.13$, respectively. However, contrary to the case of RGGR we observe that the constrained values of the model parameters differ between the flat and Gaussian prior cases. In particular, for the Gaussian prior case the $\gamma_d$ is not rigorously contained in the preferred range of $0.3-0.8$ as the Gaussian distribution allows values outside this range. Hence, we consider the flat prior results over the Gaussian ones.  In Fig.\ref{Fig:early} for each galaxy, the radial variation of the circular velocity in the DM model obtained by substituting the best-fit value is shown via the red dashed line. Similarly, the obtained parameters and goodness of fit for the other early-type galaxies in the panel for both RGGR and NFW model can be found in Table.\ref{par_sparc}. The measured goodness of fit for the early-type galaxies is compiled in Table.\ref{par_sparc} points to a positive inclination towards the RGGR scenario. 
\begin{figure*}[t]
    \centering
\begin{subfigure}{0.45\textwidth}
\includegraphics[width=\linewidth,height=0.9\linewidth]{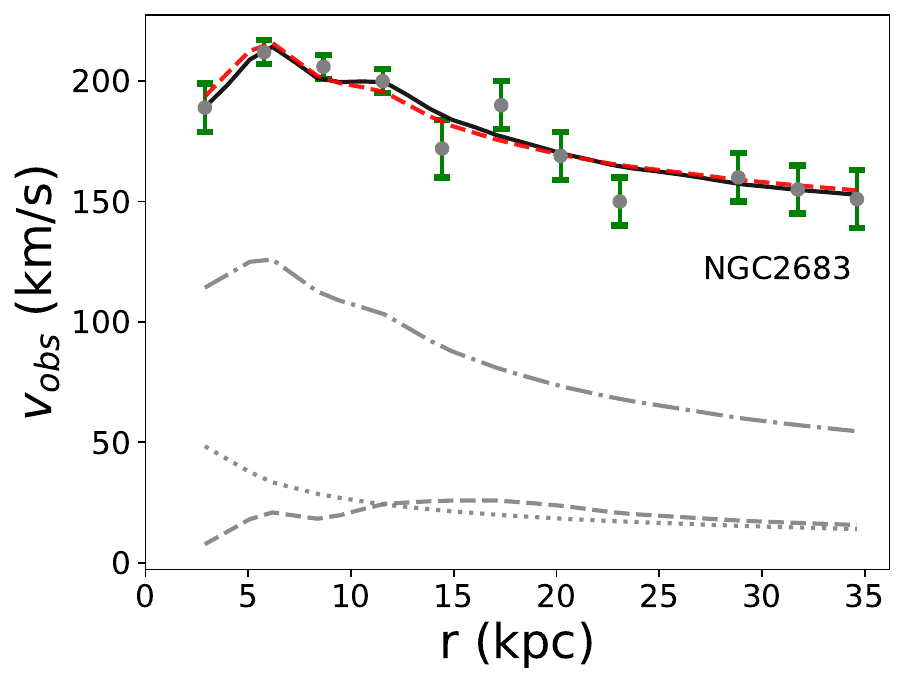} 

\end{subfigure}
\begin{subfigure}{0.45\textwidth}
\includegraphics[width=\linewidth,height=0.9\linewidth]{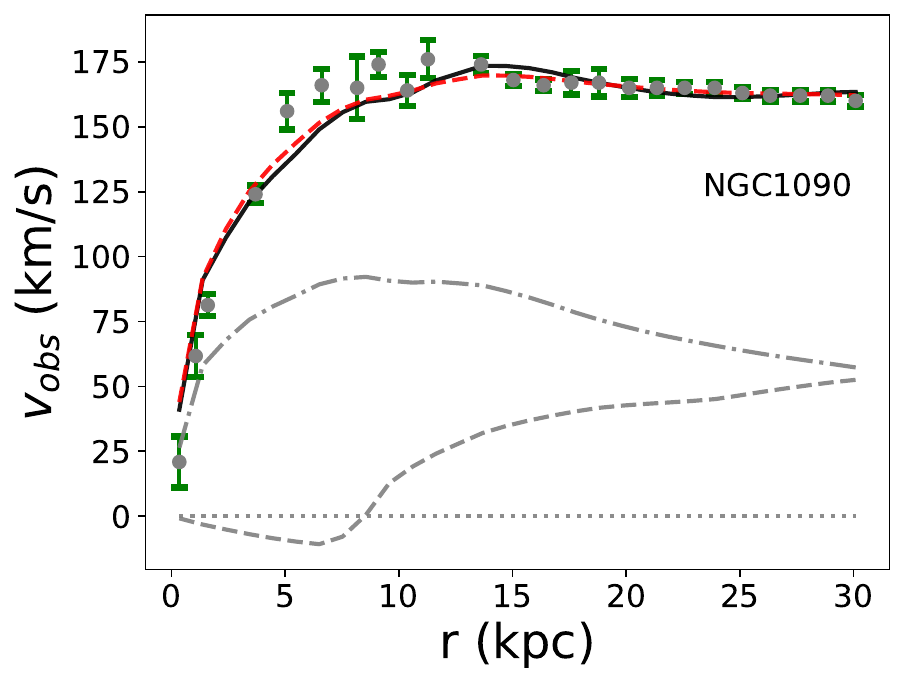} 

\end{subfigure}
\begin{subfigure}{0.45\textwidth}
\includegraphics[width=\linewidth,height=0.9\linewidth]{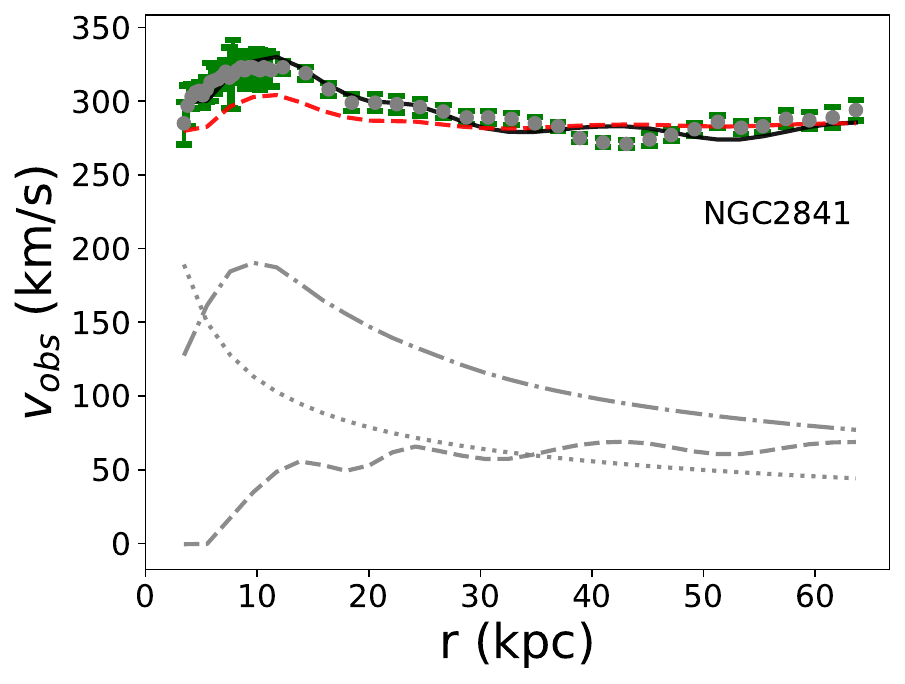} 
\end{subfigure}
\begin{subfigure}{0.45\textwidth}
\includegraphics[width=\linewidth,height=0.9\linewidth]{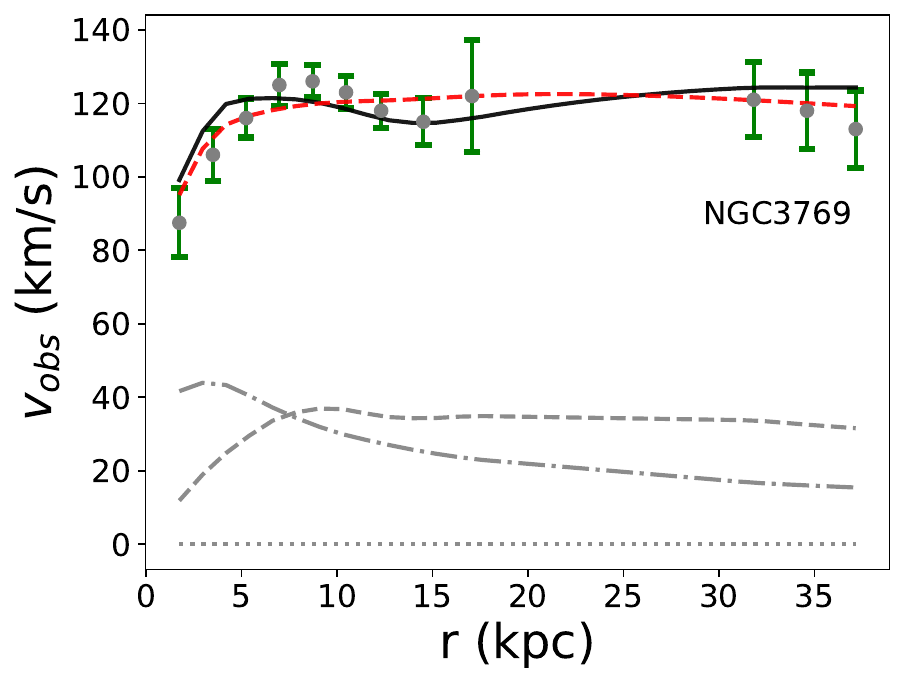} 
\end{subfigure}
\caption{For Spiral-type galaxies, the RC for $4$ specimen galaxies are shown in the panel. The black solid line is the circular velocity when RGGR gravity contributes in addition to the baryonic part. The red dash line belongs to the case where the net velocity contribution is evaluated in an NFW paradigm. The black dots with green error bars are the circular velocity data obtained from SPARC \cite{Lelli:2016zqa}. The dashed grey line represents the gaseous component of the galaxy. The dot-dashed line shows the variation of disk velocity within the galaxy, and the bulge part is plotted via the dotted line.}
\label{Fig:spiral}
\end{figure*}

\paragraph{\textbf{Spiral-type galaxies}}
\hfill\break
The second class of galaxies mentioned in SPARC belongs to the Spiral type. One example of such a galaxy is our own Milky Way. Such galaxies are characterized by the presence of distinct spiral structures at the outer parts of the disk. Additionally, the bulge at the center is of comparatively smaller size. SPARC has a large collection of spiral galaxies; our selection criteria gives us 39 such galaxies. Similar to Fig.\ref{Fig:early} we show $4$ representative galaxies i.e., NGC1090, NGC2683, NGC2841 and NGC3769 in Fig.\ref{Fig:spiral}. Out of the $4$ galaxies shown, only two have a visible bulge component present and are represented by a dotted gray line in the plot. The gray points with error bars represent the total circular velocity observed in SPARC. Similarly, the gray lines represent the individual baryonic components in a galaxy.\\
\newline
 The best-fit parameters for the first subplot for NGC2683 shown at the top left panel in Fig.\ref{Fig:spiral}. For the RGGR model when priors are assumed to be flat the best values come out to be $\gamma_d = 0.56$, $\gamma_b = 0.54$ and $~\bar{\nu} \times 10^7 = 1.88$ with $\chi^2_{red}$ equal to $0.69$. Similar to the case of Early-type galaxies, the choice of priors yields little difference in the best-fit values evaluated. In the case of NGC2683, Gaussian priors results in $\gamma_d = 0.58$, $\gamma_b = 0.50$, $\bar{\nu} \times 10^7 = 1.82$ with $\chi^2_{red}$ equal to $0.71$. For the obtained parameters, the net contribution to the total circular velocity, which is the sum of baryonic and RGGR components, is shown by a solid black line. The goodness of fits evaluated for each galaxy as given in Table.\ref{par_sparc} show that the RGGR model is a consistent choice in explaining the observed circular velocity of the spiral SPARC galaxies. Similar to early-type, RC analysis of the $39$ spiral galaxies in the presence of DM with the NFW profile is also looked into. The best-fit value for $\gamma_d$, $\gamma_b$, and $M_{200}\times 10^{11}M_{\odot}$ evaluated in the case of flat-priors for circular velocity with NFW halo results in ($0.7,0.36, 4.18$) with $\chi^2_{red}$ equals $0.84$. However, similar to the case of Early-type we consider the flat prior results out of the two choices of priors. The radial variation of circular velocity in the NFW DM model obtained by substituting the best-fit value is shown via a red dashed line for all the $4$ galaxies in the Fig.\ref{Fig:spiral}.
\begin{figure*}
  \centering
\begin{subfigure}{0.45\textwidth}
\includegraphics[width=\linewidth,height=0.9\linewidth]{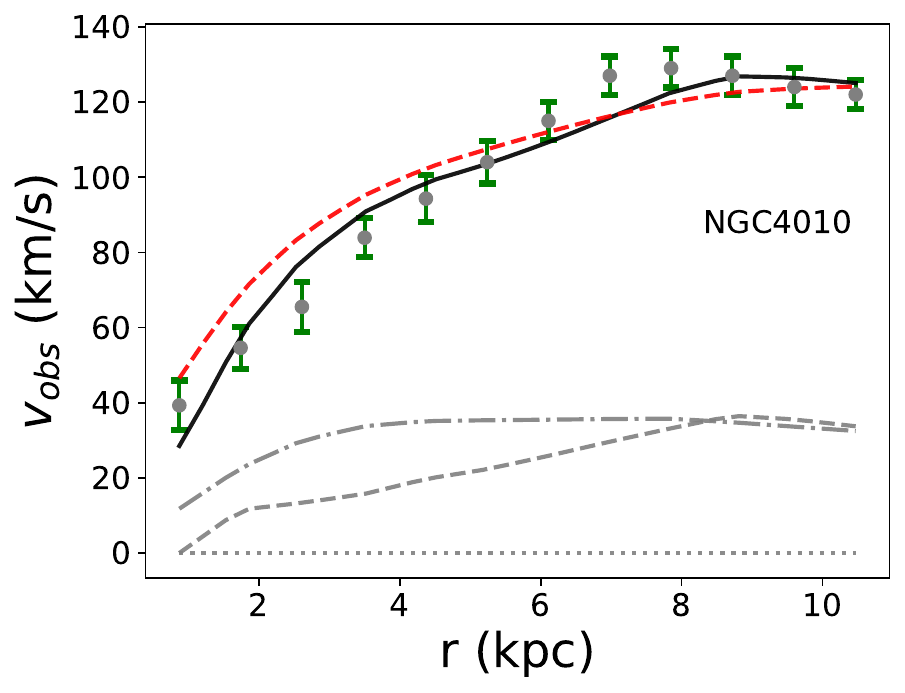} 
\end{subfigure}
\begin{subfigure}{0.45\textwidth}
\includegraphics[width=\linewidth,height=0.9\linewidth]{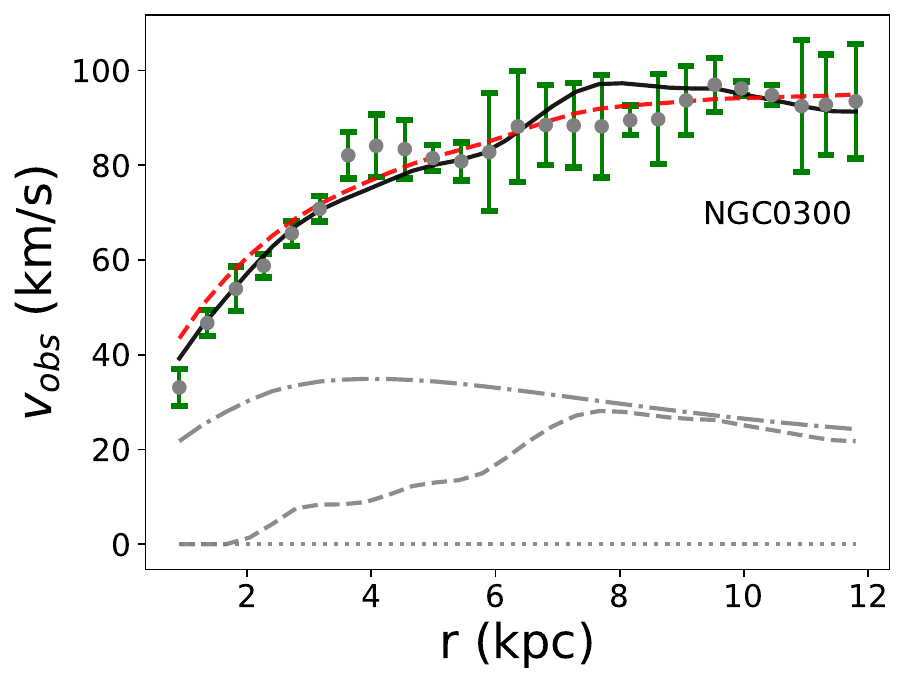} 
\end{subfigure}
\begin{subfigure}{0.45\textwidth}
\includegraphics[width=\linewidth,height=0.9\linewidth]{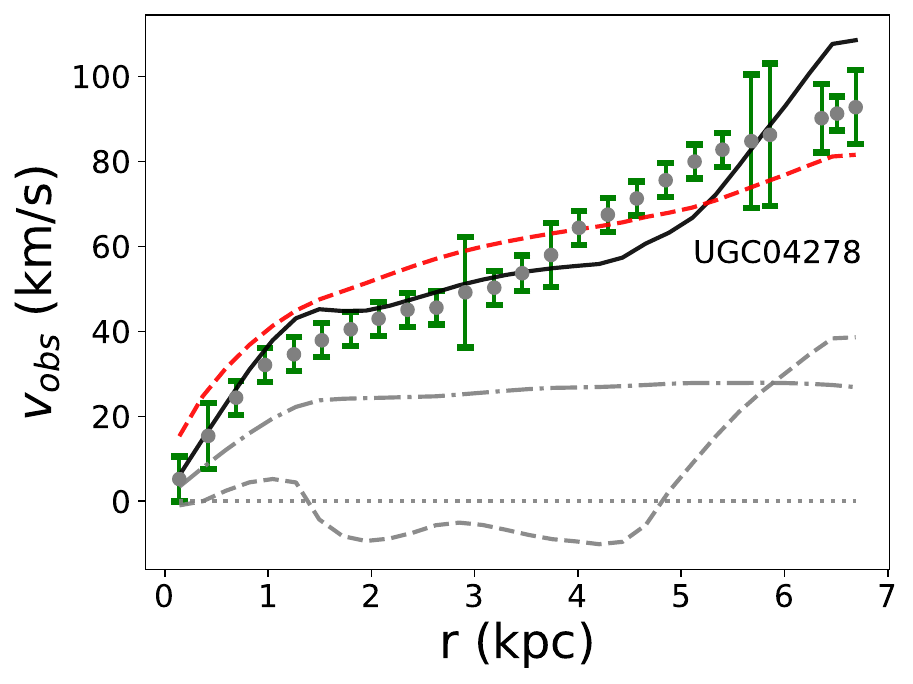} 
\end{subfigure}
\begin{subfigure}{0.45\textwidth}
\includegraphics[width=\linewidth,height=0.9\linewidth]{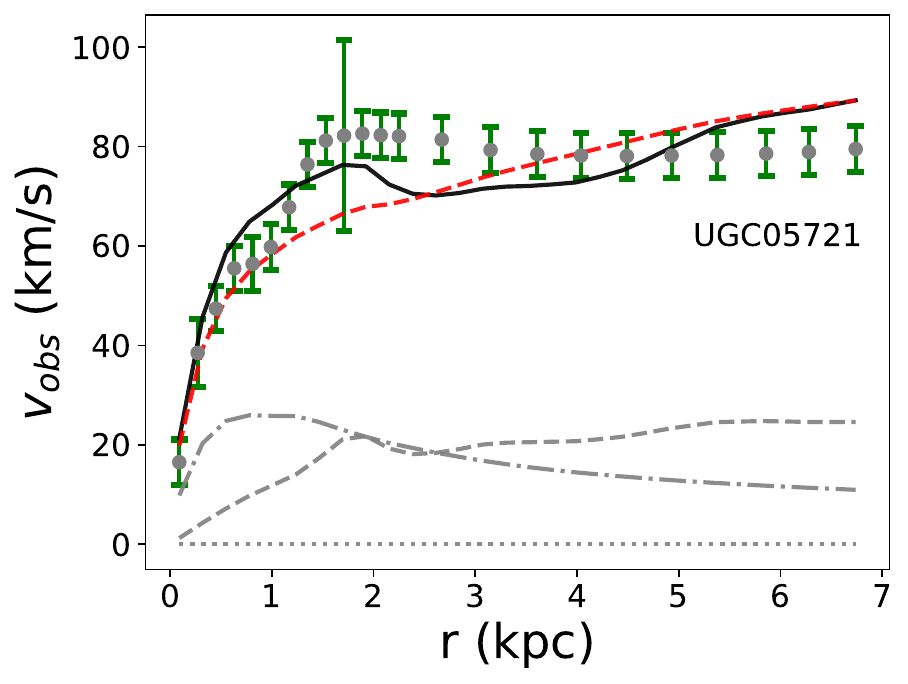} 
\end{subfigure}
\caption{For Late-type galaxies, the RC for $4$ specimen galaxies are shown in the panel. The black solid line is the circular velocity when RGGR gravity contributes in addition to the baryonic part. The red dash line belongs to the case where the net velocity contribution is evaluated in an NFW paradigm. The black dots with green error bars are the circular velocity data obtained from SPARC \cite{Lelli:2016zqa}. The dashed grey line represents the gaseous component of the galaxy. The dot-dashed line shows the variation of disk velocity within the galaxy, and the bulge part is plotted via the dotted line. }
\label{Fig.late}
\end{figure*}
\paragraph{\textbf{Late-type dwarf}}
\hfill\break
The third category of galaxies belongs to late-type dwarf, where the bulge is too faint to be observed. Thus, the baryonic component of the galaxy mostly constitutes disk and gas only. From the total late-type dwarf galaxies in the catalog, $34$ galaxies fulfill the selection criteria. Similar to the above two categories, we plot four late-type dwarf galaxies in Fig.\ref{Fig.late}. The galaxies that are illustrated include NGC4010 (top-left), NGC0300 (top-right), UGC04278 (bottom-left), and UGC05721 (bottom-right). The details on the plot corresponding to the observed circular velocity of a galaxy and its baryonic components follow the same convention as mentioned in the discussion of the previous two types of galaxies.
\newline
The circular velocity in the RGGR framework can be obtained by substituting the best-fit values of the model parameters, i.e., ($\gamma_d$, $\gamma_b$, $\bar{\nu}\times10^7$) as given in Table.\ref{par_sparc} in Eq.\ref{vrgr}. This radial variation is shown using the solid black line in Fig.\ref{Fig:spiral}. Similar to the above two morphological types, we observe a little variation in the choice of priors. In Table.\ref{par_sparc}, we report the result for the flat priors. Similarly, we also analyze the $34$ late-type galaxies in the NFW framework. The best-fit parameters obtained show preference towards the flat choice of priors and are reported along with the RGGR model in Table.\ref{par_sparc}. The $\chi_{red}^2$ computed in Table.\ref{par_sparc} for all the galaxies shows that the late-type dwarfs fit well with the RGGR model. This consistent behavior can also be visualized from the four sample plots shown in Fig.\ref{Fig.late}. 
\begin{figure*}[t]
    \centering
\begin{subfigure}{0.45\textwidth}
\includegraphics[width=\linewidth,height=0.9\linewidth]{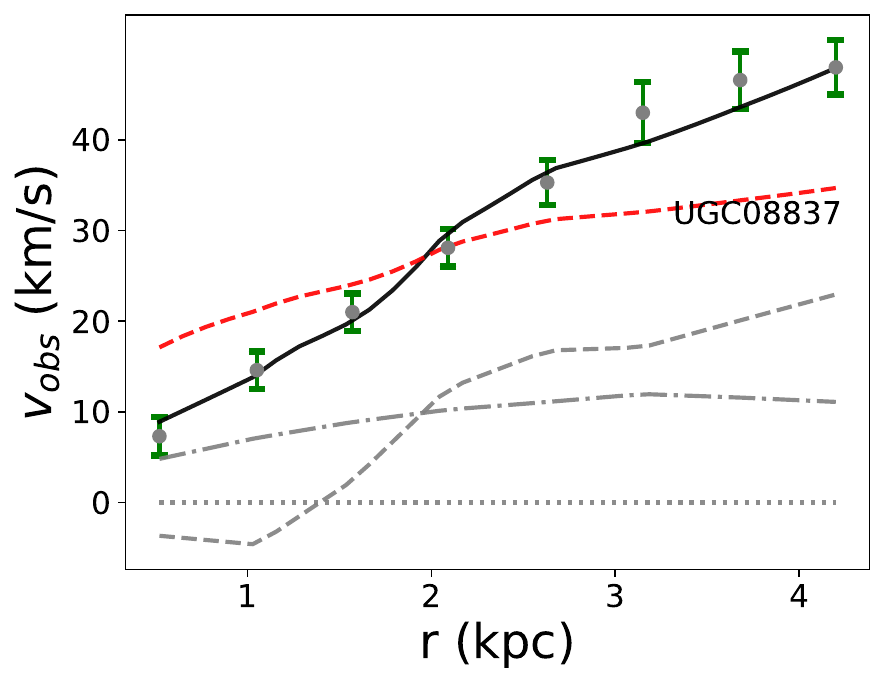}
\end{subfigure}
\begin{subfigure}{0.45\textwidth}
\includegraphics[width=\linewidth,height=0.9\linewidth]{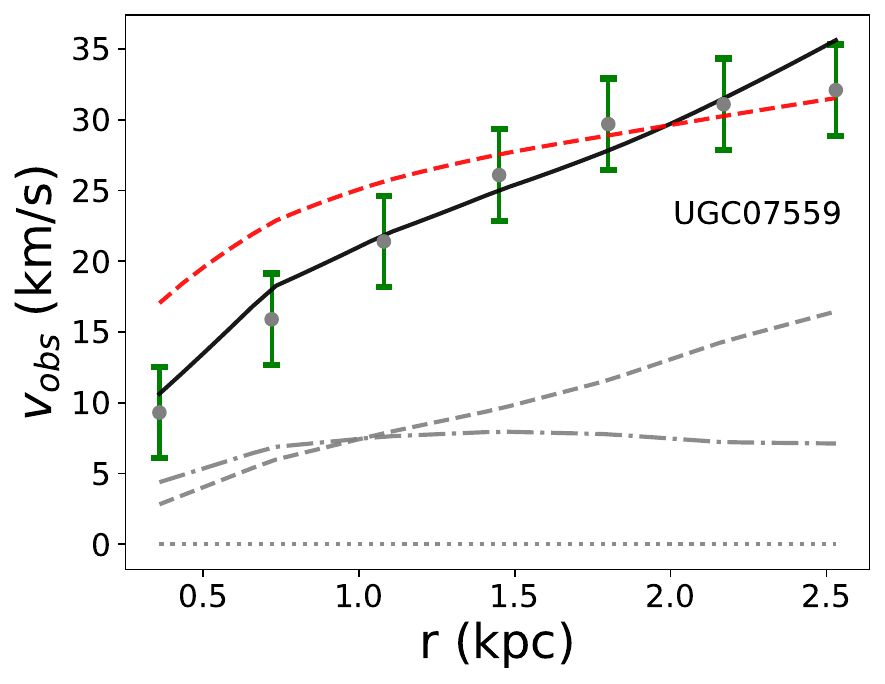} 

\end{subfigure}
\begin{subfigure}{0.45\textwidth}
\includegraphics[width=\linewidth,height=\linewidth,height=0.9\linewidth]{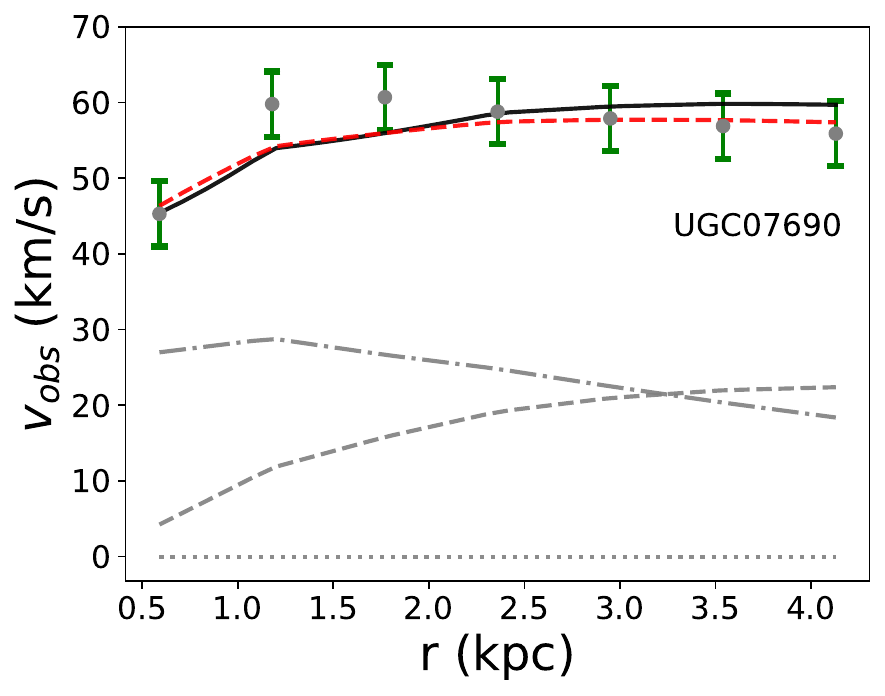} 
\end{subfigure}
\begin{subfigure}{0.45\textwidth}
\includegraphics[width=\linewidth,height=0.9\linewidth]{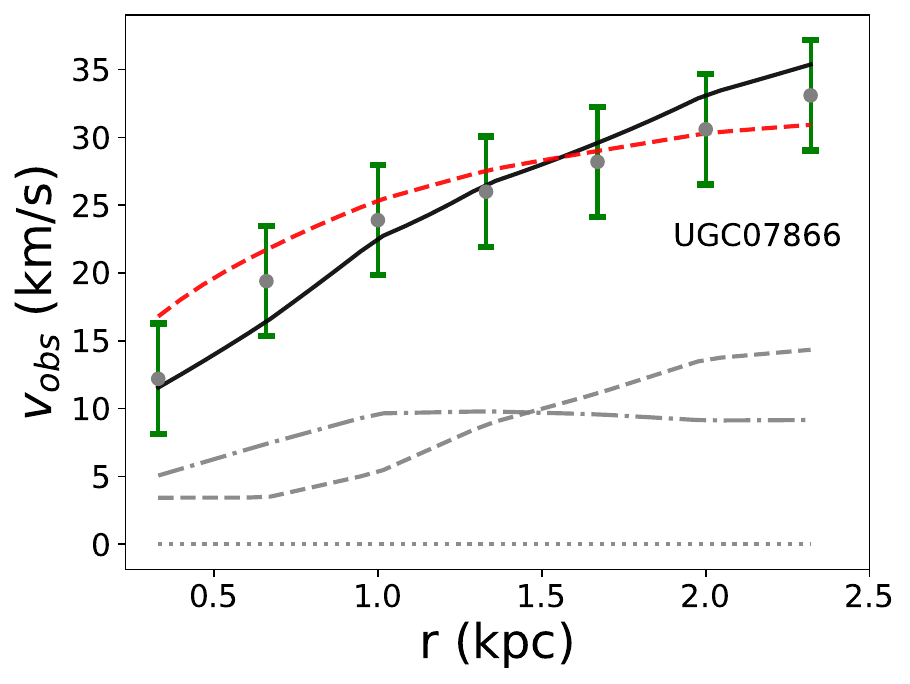} 
\end{subfigure}
\caption{For Starburst galaxies, the RC for $4$ specimen galaxies are shown in the panel. The black solid line is the circular velocity when RGGR gravity contributes in addition to the baryonic part. The red dash line belongs to the case where the net velocity contribution is evaluated in an NFW paradigm. The black dots with green error bars are the circular velocity data obtained from SPARC \cite{Lelli:2016zqa}. The dashed grey line represents the gaseous component of the galaxy. The dot-dashed line shows the variation of disk velocity within the galaxy, and the bulge part is plotted via the dotted line.}
\label{Fig:star}
\end{figure*}
\paragraph{\textbf{Starburst galaxy}}
\hfill\break

The fourth morphological type belongs to starburst galaxies which are relatively young with a high star formation rate. Such galaxies are dominated by gas and thus make a versatile region to look for the signature of the alternative gravity model. The morphology of such systems shows a diffused structure with no visible bulge and spiral arms. We studied $18$ galaxies belonging to this particular morphological type and found that RGGR can give consistent fits with the observed circular velocities. As an illustration following the convention used in other morphological types, we plot 4 starburst galaxies i.e., UGC08837, UGC07559, UGC07690, and UGC07866 in Fig.\ref{Fig:star}. The baryonic contribution for these galaxies comes from the gas and disk component only, as starbursts have no visible bulge within. This is represented in the above plot by gray lines following the same convention used in preceding cases. Also, the total observed circular velocity with error bars is shown using gray points.\\
\newline
As mentioned in the previous three morphological types, for the RGGR model, we see a little variation in our two choices of priors, i.e., flat and Gaussian.
The constrained free parameters ($\gamma_d$, $\gamma_b$, $\bar{\nu}~ \times~ 10^{7}$) obtained for the aforementioned galaxies in reference to the flat priors can be referred from Table.\ref{par_sparc}. From the obtained values of parameters, the total circular velocity evaluated for each galaxy is shown by a solid black line in Fig.\ref{Fig:star}. As can be seen from the plots, the observational circular velocity fits well with the analytical velocity in all $4$ cases. Similarly, this consistent nature of the gravity model can also be quantified for all the starburst galaxies from the goodness of fit as mentioned in Table.\ref{par_sparc}. We also studied the $18$ starburst galaxies concerning the NFW model for both flat and Gaussian priors and considered the flat prior results in preference to the Gaussian ones. The best-fit values along with the $\chi^2_{red}$ for flat-priors are compiled in Table.\ref{par_sparc}. 
For the four different morphological types of galaxies discussed above, we find that RGGR is a phenomenologically consistent theory of gravity. The best fit values for the $\bar\nu$ parameter lies in the range $10^{-6}-10^{-8}$. The parameters $\gamma_d$ and $\gamma_b$ fall in the range $0.3-0.8$, consistent with the choices of our priors. We also observe from Table.\ref{par_sparc} that the values of $\bar\nu$ are larger (in the range close to $10^{-6}$) for the early-type galaxies compared to the other types. Interestingly, the $\bar\nu$ values decrease as we go across the Hubble type towards the late-type galaxies with $\bar\nu$ around $10^{-8}$. This relation between the parameter $\bar\nu$ and the galaxy type can be understood from the baryonic mass content of the galaxies and is discussed in the next subsection.  

\begin{figure*}[t]
    \centering
    \begin{subfigure}[b]{0.45\textwidth}
        \includegraphics[width=\textwidth,height=.9\textwidth]{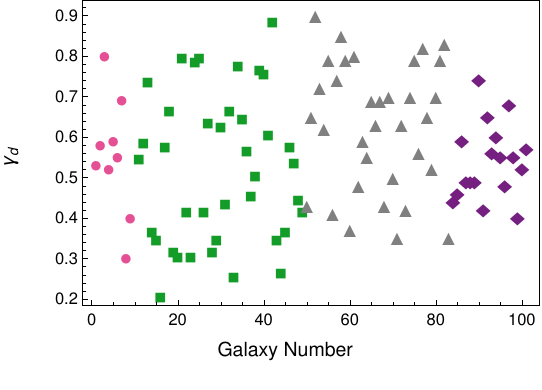}
    \end{subfigure}
    \begin{subfigure}[b]{0.45\textwidth}
        \includegraphics[width=\textwidth,height=.9\textwidth]{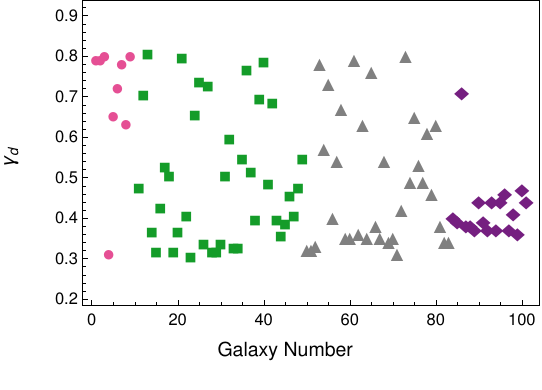}
    \end{subfigure}
    \caption{The plot shows the variation of $\gamma_d$. The left panel represents the $\gamma_d$ obtained for the RGGR analysis, and the right panel belongs to the case of NFW. The varying morphological types are represented using different markers. The square markers represent the early type, and the circles are for the spiral galaxies. Additionally, $\gamma_d$ for late-type and starburst are shown via triangle and diamond markers.}
    \label{Fig:gama}
\end{figure*}

\begin{figure}
    \centering
    \includegraphics[width=.99\linewidth, height=.4\textwidth]{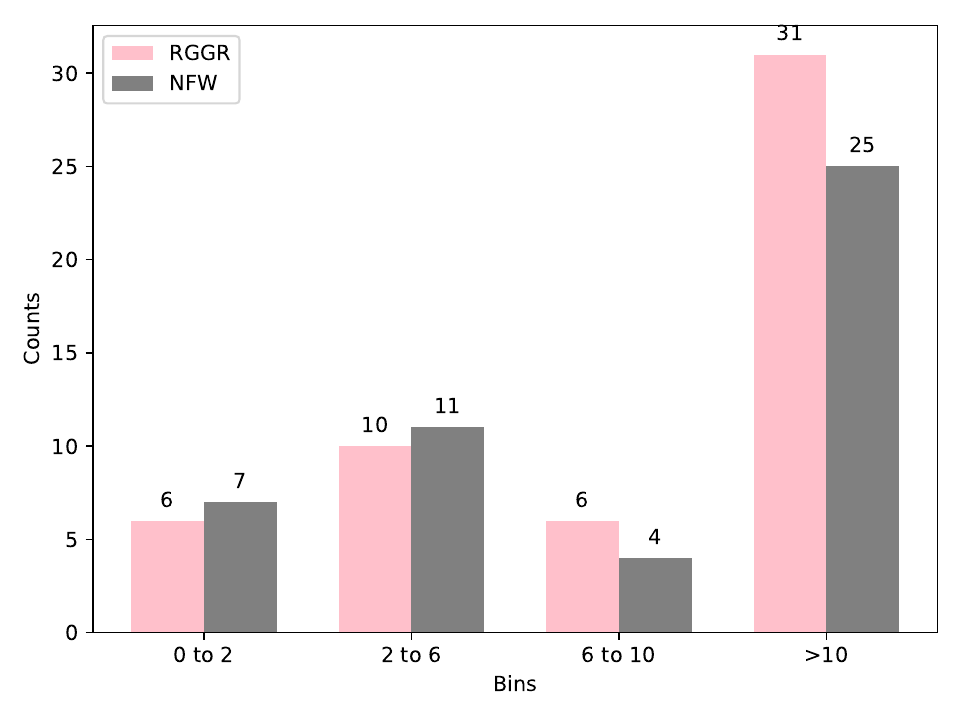}
    \caption{The plot represents the $\Delta BIC$ comparing RGGR with NFW. The pink histograms show the frequency of galaxies that favor RGGR over the DM model over different bin sizes. For $\Delta BIC$ range within $0-2$, the claim over a preference of a model is inconclusive. In total, we observe that RGGR is favored by the majority of galaxies when compared with NFW.}
    \label{Fig:bicrel}
\end{figure}

To summarise the comparison between the RGGR and the DM NFW scenario, we study $100$ galaxies that satisfy our selection criteria, as discussed previously. To measure the evidence against or in favor of our choice of models (NFW or RGGR), we compute $\Delta BIC$ for each galaxy. The plot, which measures the number of galaxies belonging to different bin sizes of $\Delta BIC$ favoring either RGGR or NFW, is shown in Fig.\ref{Fig:bicrel}. It clearly shows that out of $100$ galaxies, $47$ favors the RGGR model, and the result is inconclusive for $13$ galaxies. However, the remaining $40$ galaxies prefer the choice of DM profile. \\
 The mass modeling parameters constrained for each galaxy in our analysis include $\gamma_d$ and $\gamma_b$, which are assumed to be a constant having no radial dependence. The bulge component is prominent and can be majorly seen only in the early-type galaxies. Therefore, we utilize the disk normalization factor to show the variation of $\gamma_d$ for the galaxies of different morphological types. The plot containing the values of $\gamma_d$ for the $100$ galaxies present in our analysis is shown in Fig.\ref{Fig:gama}. For RGGR (left-panel), the behavior of $\gamma_d$ represents that this parameter varies throughout the range of $[0.3,0.8]$ for all galaxies. Also, in the NFW regime, we observe that $\gamma_d$ varies throughout the range for spiral and late-type galaxies. However, for early-type galaxies, $\gamma_d$ is concentrated towards the higher end of the prior range. Also, in the case of Starburst, we observe that $\gamma_d$ is pointing toward the lower end of the range. The different markers on the plot in Fig.\ref{Fig:gama} belong to the different morphological types of galaxies in SPARC. The pink points give the value of $\gamma_d$ for early-type galaxies.
Similarly, green squares, grey triangles, and violet diamond shapes represent $\gamma_d$ for the spiral, late-type, and starburst galaxies. Additionally, looking into the behavior of $\bar{\nu}$ in Table.\ref{par_sparc} shows that as we go from early type to starburst galaxies, the magnitude of the phenomenological parameter decreases. Such behavior of $\bar{\nu}$ with the stellar mass of galaxies is studied in detail in the next subsection.

\begin{figure}
    \centering
    \includegraphics[width=.95\linewidth, height=.4\textwidth]{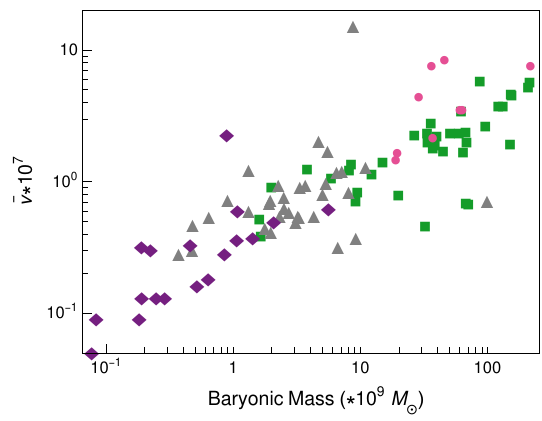}
    \caption{The plot relates the phenomenological parameter $\bar{\nu}$ with the baryonic mass of the galaxy. The individual points are the representation of the $\bar{\nu}$ evaluated for each galaxy studied in SPARC. The almost linear relation shown in the plot affirms the previous mass dependence claim for the RGGR parameter. The different markers in the plot show the different morphological types of galaxies, with the circle representing early-type, a square belonging to spiral, a triangle showing late-type dwarfs and a diamond for starburst galaxies.}
    \label{Fig:massrel}
\end{figure}
\subsection{Relation of \texorpdfstring{$\bar{\nu}$}{TEXT} with baryonic matter}
To investigate the RGGR model parameter dependence on the baryonic mass, we rely on the same galaxies from the SPARC catalog. The baryonic mass of a rotationally supported galaxy consists of the bulge, disk, and gas components, as mentioned earlier. Each component of the galaxy (stellar$+$gas) is assumed to follow a distinct density distribution. The radial variation of matter distribution for disk and gas is assumed to vary exponentially \cite{Lelli:2016zqa, Freeman:1970mx}. Similarly, for the galaxies having bulges, the density variation can be fitted with the Hernquist profile \cite{Hernquist1990AnAM}. The total baryonic content for a galaxy is a linear mass sum of stellar and gas components. The galaxies in SAPRC constitute a wide range of masses varying from $10^8-10^{11}~ M_{\odot}$ \cite{Lelli:2016zqa}.\\
\newline
For the scenario where the underlying gravity is assumed to be RGGR, the velocity contribution is the sum of the Newtonian part plus an additional term dependent on a free parameter ($\bar{\nu}$). This phenomenological parameter varies independently for each galaxy and is constrained from the study of RC as discussed in the previous section. A comparison of the constrained $\bar{\nu}$ obtained along with the stellar mass of the galaxy is represented via an individual datapoint in Fig.\ref{Fig:massrel}. The figure shows that the magnitude of the phenomenological parameter $\bar{\nu}$ increases almost linearly with the increasing mass of the galaxies. Our analysis is thus justified with the previous study \cite{Rodrigues:2014xka} done for the same gravity model. Different representations are used in the plot to categorize galaxies belonging to various morphological types. The pink round pointers belong to the early type, the green square shows the spiral, the gray triangle is for the late type, and the purple diamond represents the starburst galaxies. The figure shows that the baryonic mass decreases from the early-type galaxies to the late-type ones. Accordingly, the value of $\bar \nu$ reduces linearly: a feature which can also be seen from the data provided in Table.\ref{par_sparc}, i.e. $\bar\nu$ is smaller for the relatively younger galaxies. Such a linear relation of the model parameter with the mass of the system points towards the non-fundamental behavior of the RGGR model. However, as can be seen from the analysis of RC, the gravity model is consistent with the observations on astrophysical scales.
\subsection{Empirical relations for SPARC}
\FloatBarrier
\subsubsection{Radial Acceleration Relation (RAR)}
The data points of the SPARC galaxies projecting the relation between the observed Eq.\ref{acc} and baryonic acceleration Eq.\ref{bar} is shown to follow an empirical relation as mentioned in Eq.\ref{rar}. This shows that in weak-gravity regions, the observed and baryonic acceleration follows the relation $a_{obs}\propto \sqrt{a_{bar}}$ and depends linearly in high acceleration scales as is expressed by the empirical relation. Thus, if RGGR is a consistent gravity model, it must satisfy RAR for the SPARC galaxies. This relation has been looked into and verified for a large sample of galaxies present in the SPARC catalog without any prior assumptions for the DM or alternative gravity model. In the alternative gravity scenario, total circular velocity and acceleration are modified according to Eq.\ref{vrgr} and Eq.\ref{arggr} respectively. The net acceleration is the sum of contribution coming from different baryonic components of a galaxy ($a_{bar}(r)$) with an additional part dependent on $\phi_N(r)$ and $\bar{\nu}$ as expressed in Eq.\ref{arggr}. In our analysis, we compare the modified acceleration $(a_{RGGR}(r))$ with $a_{bar}(r)$ to study the behavior of RAR in an alternative gravity framework. \\
\begin{figure}[t!]
    
    \includegraphics[width=.94\linewidth, height=.4\textwidth]{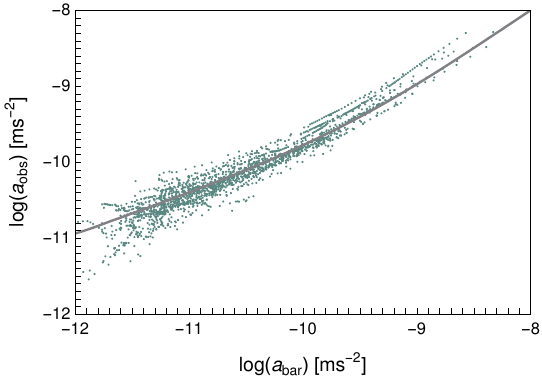}
    \caption{The plot compares the RAR relation where the total observed acceleration is evaluated in the RGGR framework. The black solid line shows the empirical relation Eq.(\ref{rar}) obtained from the SPARC observations. The green scatter points refer to the individual data points of all the galaxies analyzed where the net contribution to the acceleration gets modified as given by Eq.(\ref{arggr}) }
    \label{Fig:rar}
\end{figure}
The evaluation of the empirical relation requires the knowledge of free parameters, which is taken from the RC analysis in Sec.\ref{sec:rc}. The RAR behavior as determined for the individual datapoint of all the selected galaxies in the RGGR framework is shown in Fig.\ref{Fig:rar}. The gray solid line is the analytical relation fitted from the observational data and has the form as given in Eq.\ref{arggr}. Also, for each galaxy in our analysis, we determine the baryonic acceleration ($a_{bar}(r)$) at every radial point specified in the SAPRC catalog. Similarly, we also compute the RGGR acceleration ($a_{RGGR}(r)$) at every radial point. Thus, each green dot on the plot refers to the RGGR and baryonic acceleration at a specific radius for a given galaxy. The collection of points in Fig.\ref{Fig:rar} contains the contribution for all the 100 galaxies selected in our analysis.

Residual computed from the comparison of the relation with the 1840 data points, as illustrated in the figure, turns out to be $0.33$ dex. Thus, from the best-fit parameters from the RC analysis, RAR behaves satisfactorily in the context of the RGGR model. 
\begin{figure}[t!]
    \includegraphics[width=0.96\linewidth,height=8.0cm]{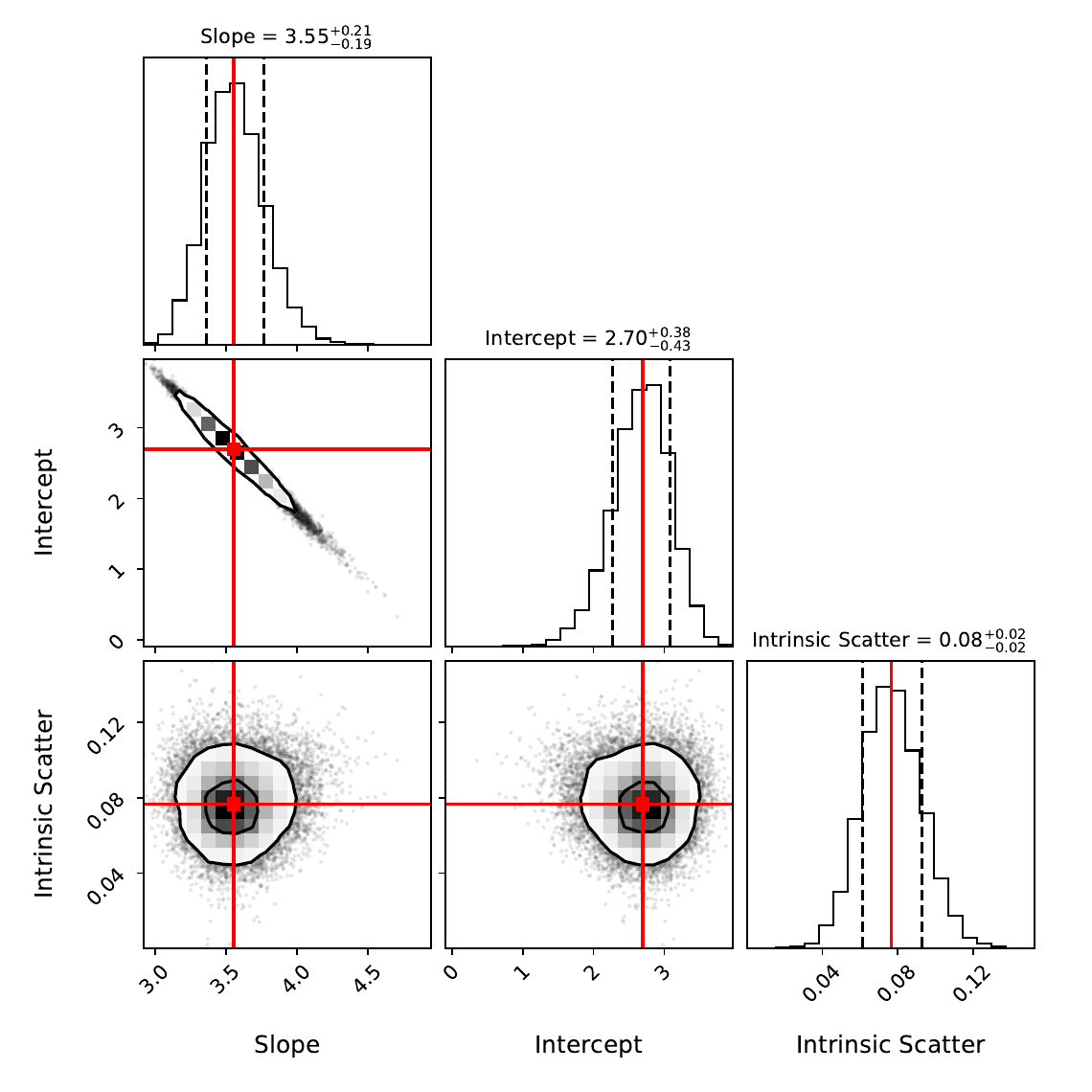}
    \caption{The posterior distribution of the parameter space in BTFR Eq.\ref{btfr_lin} evaluated using BayesLineFit \cite{Lelli:2019igz}.}
    \label{fig:btfrplot2}
\end{figure}
\subsubsection{Baryonic Tully Fisher Relation (BTFR)}
SPARC galaxies are also found to follow the BTFR (Eq.\ref{btfr}), which provides a power law relation between the baryonic mass and the velocity measured at the flat part of the rotation curve. This relation holds irrespective of the gravity model. Therefore, for RGGR to be a consistent choice as an alternative gravity model, BTFR must be satisfied by the RGGR-predicted dynamics for the SPARC galaxies. To probe the BTFR,
\begin{equation}\label{btfr_lin}
    \log(M_{bar})=x\log(V_f)+\log~A,
\end{equation}
we consider the $V_f$ corresponding to the RC fit with RGGR for each galaxy. The flat velocity ($V_f$) for each galaxy in the RGGR framework can be evaluated from the fit of the RC (Sec. \ref{sec:rc}) at a given radius. The baryonic mass $M_{bar}$ for a galaxy is a sum of stellar and gas components. The velocities of these different components are connected to their respective mass profile. For a given mass profile, $M_{bar}$ can be estimated, particularly for the SPARC galaxies, the bulge and disk are assumed to have spheroidal and exponential distribution respectively \cite{Lelli:2016zqa}. BTFR has also been found to be sensitive on the radial choice \cite{Lelli:2019igz} at which $V_f$ is measured for SPARC galaxies. We evaluate the relation at $3.2~R_d$, which has shown an optimal behavior to the relation. The choice of $3.2~R_d$ \cite{ Lelli:2019igz,romanowsky2012angular} ensures that $80\%$ of the stellar matter is encompassed within the radius. Almost for all the SPARC galaxies we select, the distance of $3.2~R_d$ from the center of the galaxy corresponds to the flat region of the RC. 
\begin{figure}[t!]
    \centering
    \includegraphics[width=0.95\linewidth,height=8.0cm]{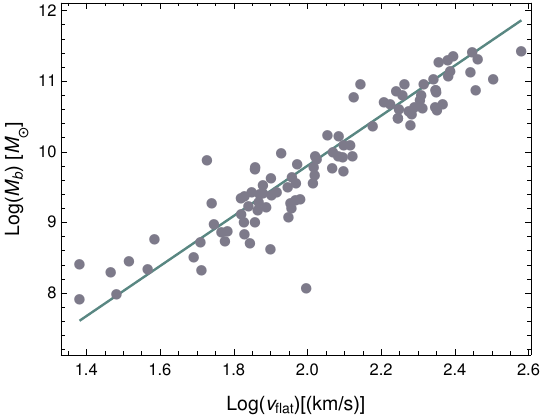}
    \caption{The BTFR with the flat circular velocity evaluated in the RGGR model. The above plot estimates $V_{f}$ for the RGGR model at $3.2~R_d$. The gray data points are the measure of flat velocity evaluated for individual galaxies. The green solid line shows the analytical best fit, consistent with the observed $V_f$ in the RGGR framework.}
    \label{fig:btfrplot1}
\end{figure}
We use BayesLineFit package \cite{Lelli:2019igz}, which is based on \textit{emcee} algorithm to find ($x$, $\log A$). The best-fit value obtained for the parameters ($x$, $\log A$) is found to be ($3.55$, $2.70$) with an orthogonal scatter of $0.08$ dex. We also report the vertical scatter given by \cite{Lelli:2019igz}:
\begin{equation}
   \sigma_0=\sqrt{\frac{1}{n} \sum^n [\log(M_{bar})-x\log(V_f)-\log A]^2},
\end{equation}
where $n$ is the number of galaxies selected in our analysis, the observed $\sigma_0$ for the case of $3.2~R_d$ comes out to be $0.54$. The posterior distribution for the fitting parameters, i.e., slope ($x$), intercept ($\log~A$), and intrinsic scatter, are shown in Fig.\ref{fig:btfrplot2}. The solid red lines in the plot point towards the maximum likelihood values for the parameters. Given the best-fit parameters, BTFR for all the galaxies included in our analysis is illustrated in Fig.\ref{fig:btfrplot1}. The solid gray circles represent the flat velocity for each SPARC galaxy evaluated at $r=3.2~R_d$. Additionally, the green solid line fits the linear equation Eq.\ref{btfr_lin} for the best-fit parameters obtained in our analysis. The BTFR plot in Fig.\ref{fig:btfrplot1} shows that the flat velocity evaluated at $3.2~R_d$ consistently matches the observational results.

\section{Conclusion}
\label{sec:conc}
Our analysis focuses on the renormalization group improved gravity to explain the kinematics for a collection of rotationally supported galaxies compiled in the SPARC catalog. 
We, for the first time, study the dependence of the alternative gravity parameter on the galaxy morphology. We have looked into four different morphological types of galaxies, viz. early, spiral, late, and starburst. We have constrained the model parameter $\bar \nu$ for each galaxy taken from all four morphological types and have found that the RGGR consistently fits the observed net circular velocity. Our statistical analysis has probed this consistency for the individual galaxies using the goodness of fit. We have also checked the linear dependence of the model parameter $\bar \nu$ on the baryonic mass of the galaxy.
The constrained values for the parameter $\bar\nu$ for our sample of the SPARC galaxies lie in the range $10^{-6}-10^{-8}$ and are consistent with the variation of the masses of the galaxies under consideration. This implies that the older and heavier galaxies lead to a larger $\bar \nu$. Indeed, we have found that the parameter $\bar\nu$ decreases from the older galaxies to the younger ones. 
\newline
We have further verified our goodness of fit from the RC analysis in light of the well-known empirical relations: RAR and BTFR.
In particular, the RAR compares the radial variation of the observed and the baryonic acceleration. An additional factor comes in the baryonic acceleration due to the RGGR model. Our analysis has found that the RAR in the RGGR framework aligns with the established analytical relation obtained from previous observations. 
On a similar note, BTFR compares the baryonic mass with the observed flat velocity ($V_f$) of a galaxy, suggesting a power law dependence between them. The exact power law index has been found to be sensitive to the choice of the flat velocity radius. We have selected a radius of $3.2 R_d$, which is known to show optimal behavior with the observations. Our analysis finds a tight correlation of the baryonic mass with flat velocity measured at $3.2R_d$ and has obtained a small orthogonal scatter of $0.08$. 
This scatter may reduce even further for different choices of the radius of the $V_f$ and by relieving the assumption of universal exponential density profile for the diffused gas component of the galaxy, which requires further study.
\\
\newline
Additionally, we compare the RGGR model with a DM scenario, where we assume the profile to be NFW. Our analysis clearly shows that, leaving aside the galaxies where both the models (RGGR and NFW) perform equally well, RGGR is favored in relatively larger number of galaxies than NFW. 
An important aspect of the RGGR model is that the model parameter $\bar\nu$ is dependent on the mass of the gravitating system. Many other alternative gravity models exhibit dependence of the model parameters on the scale of the system. Though this scale dependence of the phenomenological parameter may question the fundamental nature of these alternative gravity models, they still remain a consistent theory of gravity. It is to be noted that such RG models of gravity face certain criticisms for the non-universal choice of the running scale parameter \cite{Donoghue:2019clr}. However, for any alternative gravity model to be considered viable, it must satisfy the consistency criteria, i.e., relativistic, self-consistent, and a complete theory free of ghosts and instabilities. The theory also requires the ability to explain the gravitational dynamics both in the strong and weak-field regimes as well, and it should regain a valid Newtonian limit on the Solar System scales \cite{will_1993, Shankaranarayanan:2022wbx}. RGGR, an alternative gravity model, follows the above consistency criteria and thus remains a phenomenologically consistent theory of gravity. Most importantly, in the context of our work, RGGR satisfies the galactic kinematic observations for a large selection of SPARC galaxies. 
Further validation of the RGGR model as an alternative to the DM paradigm requires the model to be tested in expanded galactic mass scales, for example, the newly discovered massive galaxies  \cite{Labb2022APO}. In the smaller galactic mass scales, the DM-dominated ultra-diffuse galaxies like DF44 \cite{vanDokkum:2016uwg} will provide further tests \cite{eshabh} to the RGGR gravity model.
\acknowledgments
Sovan Chakraborty acknowledges the support of the funding from DST-SERB
projects CRG/2021/002961 and MTR/2021/000540. Sayan
Chakrabarti would like to acknowledge the support from
the DST-SERB research grant MTR/2022/000318. All the
authors would like to acknowledge discussions and valuable inputs from Sayak Dutta. Esha Bhatia is indebted to the Ministry of Human Resource Development, Government of India, for assistance through a doctoral fellowship.

\nocite{*}

\bibliography{apssamp}

\end{document}